\documentclass[letterpaper,11pt]{emulateapj}
\usepackage{epsfig}
\usepackage{natbib}

\begin{document}
\newcommand{\kms}{km~s$^{-1}$}
\newcommand{\Msun}{M_{\odot}}
\newcommand{\Lsun}{L_{\odot}}
\newcommand{\ML}{M_{\odot}/L_{\odot}}
\newcommand{\etal}{{et al.}\ }
\newcommand{\hhh}{h_{100}}
\newcommand{\hsq}{h_{100}^{-2}}
\newcommand{\tn}{\tablenotemark}
\newcommand{\mdot}{\dot{M}}
\newcommand{\p}{^\prime}
\newcommand{\kmsMpc}{km~s$^{-1}$~Mpc$^{-1}$}

\title{The Extragalactic Distance Database: All Digital HI Profile Catalog}

\author{H\'el\`ene M. Courtois}
\affil{Universit\'e Lyon 1, CNRS/IN2P3/INSU, Institut de Physique Nucl\'eaire, Lyon, France and Institute for Astronomy, University of Hawaii, Honolulu HI 96822}


\author{R. Brent Tully,}
\affil{Institute for Astronomy, University of Hawaii, 2680 Woodlawn Drive, Honolulu, HI 96822}


\author{J. Richard Fisher}
\affil{National Radio Astronomy Observatory\footnote{The National Radio Astronomy Observatory is a facility of the National Science Foundation, operated under cooperative agreement by Associated Universities, Inc.}, 520 Edgemont Road, Charlottesville, VA 22903}


\author{Nicolas Bonhomme}
\affil{Universit\'e Lyon 1, CNRS/IN2P3/INSU, Institut de Physique Nucl\'eaire, Lyon, France}


\author{Maximilian Zavodny}
\affil{Institute for Astronomy, University of Hawaii, 2680 Woodlawn Drive, Honolulu, HI 96822}

\and

\author{Austin Barnes}
\affil{Institute for Astronomy, University of Hawaii, 2680 Woodlawn Drive, Honolulu, HI 96822}

\begin{abstract}
An important component of the Extragalactic Distance Database (EDD) at http://edd.ifa.hawaii.edu is a group of catalogs related to the measurement of HI line profile parameters.  One of these is the {\it All Digital HI} catalog which contains an amalgam of information from new data and old.  The new data results from observations with Arecibo and Parkes telescopes and with the Green Bank Telescope (GBT), including continuing input since the award of the NRAO Cosmic Flows Large Program.  The old data has been collected from archives, wherever available, particularly the Cornell University Digital HI Archive, the Nan\c cay Telescope extragalactic HI archive, and the Australia Telescope HI archive.  The catalog currently contains information on $\sim 15,000$ profiles relating to $\sim 13,000$ galaxies.  The channel -- flux per channel files, from whatever source, are carried through a common pipeline.  The derived parameter of greatest interest is $W_{m50}$, the profile width at 50\% of the mean flux.  After appropriate adjustment, the parameter $W_{mx}$ is derived, the linewidth which statistically approximates the peak to peak maximum rotation velocity before correction for inclination, $2 V_{max} {\rm sin} i$.
\end{abstract}


\section{Introduction}

The goal of the overall program facilitated by the Extragalactic Distance Database (EDD) is to obtain the densest and deepest possible coverage of galaxy distances and, hence, of line-of-sight peculiar velocities.  We want to improve the local determination of the Hubble Constant and measure departures from the cosmic expansion that presumably can be attributed to the distribution of matter.  We are giving consideration to 7-10 different methods for deriving distances.  One of these relies on the correlation between galaxy luminosities and rotation rates \citep{1977A&A....54..661T}.  While there are other methods that provide more accurate individual measures of distance, the luminosity--rotation rate correlation retains an importance.  The method can be applied to roughly half of all spiral galaxies -- those that are suitably inclined and not confused or disrupted by companions --  over a wide range of distances.  Highly complete samples of many thousands of galaxies can be obtained extending to 40--100 Mpc, constrained only by manpower and competition for access to telescopes.

The measurement of a distance requires separate observations of luminosities and rotation rates.  The latter can be determined through spectroscopy in either optical or radio domains but care must be taken if sources are mixed \citep{1997AJ....114.2402C,2007AJ....134..334C}.  
Usually radio spectra are obtained with beams that encompass and integrate the flux over an entire target, providing a global line profile.  Information obtained with radio telescopes typically provide higher spectral resolution than optical alternatives.  Spatial coverage is almost always more extensive.  To date, optical spectra have been obtained for distance measurements over a greater range \citep{1999AJ....118.1489D}, although respectable HI spectra have been obtained with Arecibo Telescope at redshifts as great as $z \sim 0.2$ \citep{2008ApJ...685L..13C}.  One wants ultimately to reconcile optical and radio rotation curve information. This article focuses on the narrower issue of the measurement of rotation rates from radio observations of the neutral Hydrogen 21 cm line.  The purpose of the present discussion is to integrate the considerable amount of neutral Hydrogen spectral data obtained by ourselves and others through a coherent analysis.  

The study involves the following elements.  First, there is a review of the status of neutral Hydrogen observations as they pertain to our program.  Then, we discuss our own recent observations with the Arecibo, Green Bank, and Parkes telescopes.  A reduction procedure has been developed to analyze the data.  Our observations incrementally expand on the large body of material available in the archives of radio observatories around the world.  There is a discussion of the unification of all the available digital data, analyzed in a common way.  The results are provided in tabular and graphical forms in EDD, the Extragalactic Distance Database \citep{2009AJ....138..323T}.  See http://edd.ifa.hawaii.edu; select the catalog {\it All Digital HI}.

\subsection{Historical Background}

Motions within galaxies are a response to the gravitational potential.  If the HI gas is in equilibrium in a disk, rotating in circular orbits, then there is a simple relationship between the observed motions and the distribution of mass.  The small dispersion in the relation between rotation rate and luminosity implies a strong correlation between the dark matter that dominates the potential and the baryonic matter that shines.  Considerable effort has been made to try to understand this link \citep{2003MNRAS.343..367B, 2007ApJ...654...27D}.  A focus of recent attention has been on the evaluation of how the correlation changes with look-back time \citep{2005ApJ...628..160C, 2006ApJ...653.1049W}.  Our main interest is more modest: use of the empirical correlation as a way to measure distances \citep{2000ApJ...533..744T, 2008ApJ...676..184T}.

Neutral Hydrogen is easily detected in nearby spiral and irregular galaxies with modern radio telescopes.  The product of an observation with a single dish facility is a line profile which can be grossly characterized by three parameters: an integrated flux, a systemic doppler shift from the rest wavelength, and a linewidth due to internal motions.  The distances over which galaxies can be detected depends on the sensitivity of telescopes and the intrinsic gas content of galaxies.

Regarding telescopes, the bigger the better.  Consider the situation that is generally close to being met of unresolved sources.  Receivers and efficiencies being equal, the advantage of a big telescope in exposure time required to reach a given signal-to-noise goes as the fourth power of the aperture.   The signal-to-noise, $S/N$, achieved in a unit time, $t_0$, depends on the square of the aperture, $D$, of the telescope.  To reach a specific signal-to-noise requires a time $t$:  $S/N \propto (t/t_0)^2 \propto D^4$.  Arecibo Telescope is presently by far the most sensitive single dish instrument for HI line studies.  Unfortunately it only accesses 30\% of the sky.  An additional advantage of a large telescope is a relatively small primary beam, hence reduced source confusion.  A small beam only becomes a disadvantage when it is smaller than the dimensions of the source, a situation that can result in lost flux and a biased linewidth. 

\begin{figure}[htb!]
\figurenum{1}
\centering
\includegraphics[scale=0.4]{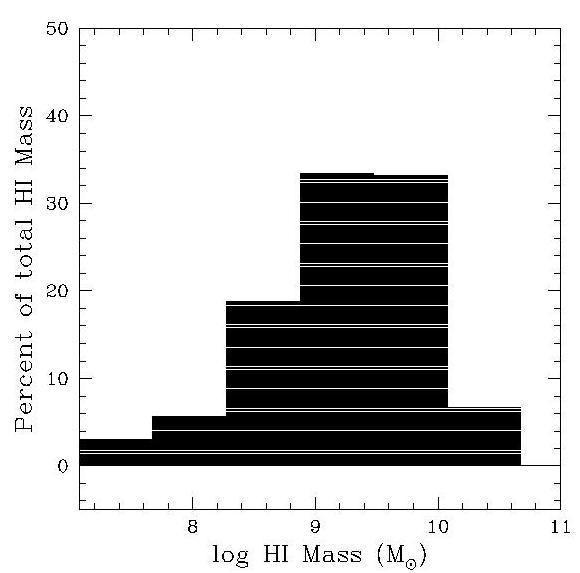}
\includegraphics[scale=0.4]{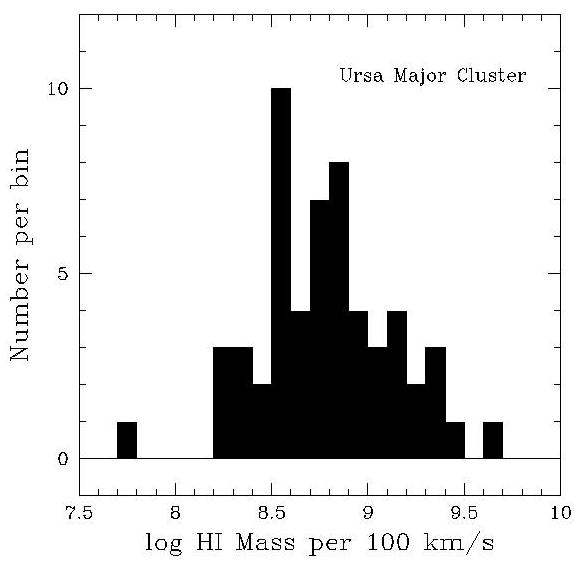}
\caption{{\it Top:} Fraction of HI mass in logarithmic mass intervals derived from the HI mass function of \citet{2005MNRAS.359L..30Z}.  {\it Bottom:} HI mass per 100 \kms\ of linewidth for a sample drawn from the Ursa Major Cluster complete to an HI mass limit of $10^7~M_{\odot}$.}  
\label{HIfrac}
\end{figure}

\begin{figure}[htb!]
\figurenum{2}
\centering
\includegraphics[scale=0.4]{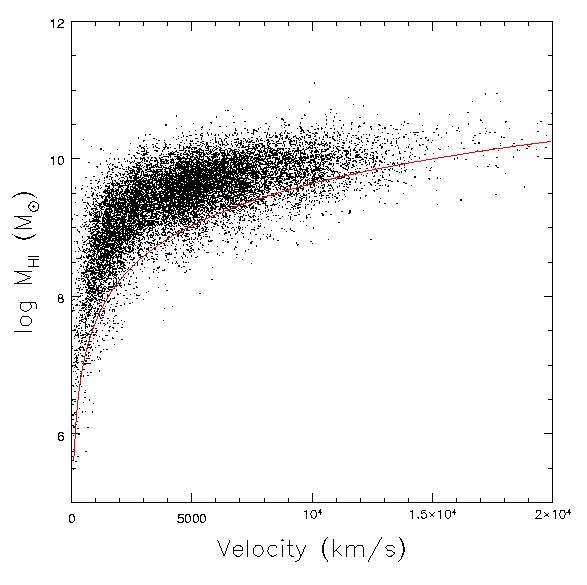}
\caption{HI mass assuming distance = $cz$ \kms / 75 \kms\ Mpc$^{-1}$ vs. systemic velocity for 15,000 galaxies in the Lyon Extragalactic Database (LEDA).  The solid curve traces the locus of an integrated flux of 1 Jy km s$^{-1}$, a practical limit with the Arecibo Telescope.}
\label{MHI_V}
\end{figure}

Regarding the properties of galaxies, it is instructive to consider the HI mass function  \citep{2005MNRAS.359L..30Z}.  The cut-off at high mass is abrupt. Systems with higher HI mass than $3 \times 10^{10}~\Msun$ are rare.  It can be supposed that larger gas reservoirs than this limit quickly get converted into stars.  By contrast, many so-called dwarf galaxies have abundant HI, reflecting low time-averaged star formation rates.  The result is a pile-up of 90\% of HI masses in the two decade range $1.5 \times 10^8 - 1.5 \times 10^{10}~\Msun$.  Figure~\ref{HIfrac} demonstrates that most of the neutral gas in the $z=0$ Universe is  locked up in galaxies (which are mostly spirals) with ${\rm log} M_{HI}/\Msun = 9.2 \pm 1.0$.  In the second panel, a histogram is presented of HI mass divided by profile linewidth (measured at 20\% of peak intensity in units of 100 \kms).  The sample is drawn from the Ursa Major Cluster at 17 Mpc and is essentially complete to an HI mass of $10^7~M_{\odot}$ \citep{1996AJ....112.2471T}.  Eighty-five percent of the sources are contained within a single decade $2 \times 10^8 - 2 \times 10^9~M_{\odot}$ per 100 \kms. Total intrinsic fluxes are constrained to a two decade window (left panel of the figure) and, since sources with greater fluxes tend to have larger linewidths, intrinsic fluxes per spectral channel are constrained to a single decade (right panel).  The curious consequence is that galaxies typed Sb--Sc--Sd can be detected in HI with comparable likelihood.  A program to observe these kinds of galaxies can be expected to have a high level of completion within a volume dictated by telescope, time available, and motivation.  Current capabilities can be evaluated by giving consideration to Figure~\ref{MHI_V}.  Note the sharp upper cut-off in $M_{HI}$ which is rather flat with distance (increased volume).

It is one thing to detect HI in a galaxy and quite another to obtain a signal useful for the determination of distances.  Non-pathological line profiles have characteristic features that are helpful.  In the case of massive galaxies, most of the flux is at the high and low frequency extremes, originating from gas on the flat part of rotation curves.  The consequence is profiles with abrupt edges and flux where it is most useful to define these edges.  For galaxies of sufficiently low mass, those with more slowly rising rotation curves, the line profiles lose the two-horn shape and instead can be approximately gaussian.  These systems tend to have less total HI but it is piled up over a small range of wavelengths, feeding the conspiracy of a common detectability level. 

\section{Existing HI Information}

There has been a transformation in the last decade in the way data is taken at radio telescopes.  In the early days, the observer made measurements from analog displays and the published output were profiles on a journal page.  The catalog in the Extragalactic Distance Database (EDD) called {\it Pre-Digital HI} gives a compilation of information laboriously extracted from literature profiles from many dozens of sources (cf. \citet{1989gcho.book.....H}).  The great concern and attention in the compilation of this historical catalog was with profile linewidths (the fluxes in the catalog are of mixed quality and may often suffer from beam dilution effects).  Attention should be given to the parameter $W_{20}$, the linewidth at 20\% of peak intensity. Often, profiles for a given galaxy are available from multiple sources.  There is a velocity cutoff of 3,000~\kms\ for the entries in this catalog.  A significant contribution came from observations by two of the authors \citep{1981ApJS...47..139F}.  It is to be emphasized that the linewidths given in the {\it Pre Digital HI} catalog do $not$ come from the literature source; they were all derived from published graph profiles by Fisher, Tully, or our colleague Cyrus Hall.

Some use will be made of this pre-digital information in what follows.  This carefully accumulated  information will be used as a basis of comparison with digital data processed by machine.  A value from the {\it Pre Digital HI} catalog can even be preferred as a primary source in some cases of very large, nearby galaxies that could only be satisfactorily observed with the large beams of small telescopes that are no longer in service.  Of course, care must be taken to avoid systematics in the translation of profile information between different systems.

The turn of events that marginalizes the {\it Pre Digital HI} material is the ubiquitous availability of digital spectral information from radio telescope archives.  There is no need to go to the telescope to access a rich load of data.  Of course, the authors of each source will have analyzed their material in their own way.  At issue for us, though, is the need to analyze all profile information from whatever source in a common way.  One less than satisfactory approach would be to ingest parameters such as line widths, fluxes, and systemic velocities from literature sources and attempt to reconcile related measurements.  However more coherent results can be expected if a standard analysis procedure is applied to the directly observed spectra.

There are very large and very good compilations of HI material already available.  The most important are assembled in separate catalogs within EDD.  Foremost, there is the extensive database compiled by the Cornell group labelled {\it Springob/Cornell HI} \citep{2005ApJS..160..149S} which presents material obtained with the Arecibo Telescope, the old 140-foot and 300-foot Green Bank telescopes, and limited material from the Nan\c cay and Effelsberg telescopes.  Additionally, EDD contains tables and links relating to observations by other observers at individual telescopes.  The results of a major observing program with Nan\c cay Telescope \citep{2006ASPC..351..429T} are accessed through the catalog {\it HI Nancay}.  Key results from the Parkes Telescope all-southern-sky multibeam project \citep{2004AJ....128...16K} are reported in the catalog {\it HIPASS 1000}.

Although the material available from archives is extensive, it is not sufficient.  There are galaxies that are important to our program that have not been satisfactorily observed.  Details regarding our samples will be described elsewhere but here is a flavor.  One sample contains all suitably inclined and unobscured spirals within 3,000~\kms\ brighter than $\sim 0.3 L^{\star}$ (see the catalog {\it V3k $M_K<-21$} in EDD).  Another is of spirals out to 6,000~\kms\ selected from the Infrared Astronomical Satellite Point Source--Redshift Catalog (IRAS PSC-z; see the catalog {\it Saunders PSCz} in EDD).  Yet another sample consists of galaxies extending out to 10,000~\kms\ with observed supernovae of Type Ia, included for the purpose of improving the zero-point calibration of the supernova scale.  

For the present discussion, the point to be made is that we needed to supplement the archival material with our own observations, which required that we develop procedures for the analysis of our observations.  We might have simply adopted a pre-existing procedure, say, incorporating one of the linewidth measures described by \citet{2005ApJS..160..149S}.  However, we decided we would learn more if we developed our own algorithm, albeit one inspired by an approach considered by Springob et al.  The exercise would provide independently derived results that could, in many cases, be compared with other sources to evaluate uncertainties.

The discussion continues in the next section with descriptions of our observations with the Arecibo, Green Bank, and Parkes telescopes.  In the ensuing four sections there are descriptions of how we measure profile parameters, evaluations of product quality, and a brief discussion of adjustments to give a parameter of dynamical interest.  In the section before the summary, there is a description of the catalog in EDD called {\it All Digital HI} which contains both the results of our own observations and the reprocessed results of data from the archives, all treated with the same procedures.

\section{Observations with the Arecibo, Green Bank, and Parkes Telescopes}

Our program began once the Arecibo Telescope was brought  back in service after installation of the Gregorian feed and ground screen.  Since the start of the multibeam sky survey Arecibo Legacy Fast ALFA (ALFALFA), we have discontinued our own observations with Arecibo Telescope, with the expectation that many of our sources in the survey range $0\degr < \delta < 36\degr$ will be observed serendipitously with sufficient accuracy.  Subsequently, we have been observing with the 100m Green Bank Telescope at declinations above $\delta = -45\degr$ but excluding the Arecibo range.  As of the third semester of 2008 this program has been awarded the status of a Large Program, now christened with the name {\it Cosmic Flows}\footnote{http://www.vla.nrao.edu/astro/prop/largeprop/}.  Results from this program are appearing in the {\it All Digital HI} catalog as they become available.  Access to the remaining sky, at $\delta < -45\degr$, requires use of Parkes Telescope in Australia.  Observations with this facility began early in 2009.

\subsection{Arecibo}

The single-beam Arecibo observations of 330 galaxies were undertaken in two sessions, October
1999 and April 2001, mostly between sunset and sunrise to avoid
spectral baseline distortion from solar continuum emission.  The 1999
session used the "L-Narrow" receiver for objects with known redshifts
and the "L-Wide" receiver for unknown redshifts that required a
search over a wide velocity range.  The 2001 session used only the
"L-Wide" receiver with improved system noise temperature.

Eight correlator sections, each with 2048 spectral channels, were
available.  The unknown redshift search used 4 overlapping, 25-MHz
bandwidth spectra for a velocity span from -400 to +18,000 \kms,
heliocentric, in each of two linear polarizations.  The known redshift
spectra were taken with 4 correlator sections, 2 polarizations with
12.5 MHz bandwidth and 2 with 6.25 MHz bandwidth, all centered on the
galaxy line profile.  All spectral data were Hanning smoothed to
produce resolutions of 1.29, 2.58, and 5.16 \kms\ at zero redshift with
6.25, 12.5, and 25 MHz bandwidths, respectively.
Narrowband RFI was edited manually and data values replaced with a linear interpolation.  For most spectra a second-order baseline was least-squares fit to the data on either side of the line profile and subtracted from all spectral data values.  In no case was the baseline curve higher that third order.

The basic Arecibo observation was 7 minutes on the object and 7
minutes on a blank-sky position on the same hour angle track.  When
time was available weak line profiles were observed on more than one
day and the spectra were averaged.
The line profile flux density scale was established with correlator
observations of known continuum sources over the full scan range of
the Arecibo telescope to determine gain as a function of zenith
distance.

\subsection{Green Bank}

The single-beam Robert C. Byrd Green Bank Telescope (GBT) observations were carried out in the course of two programs,
one program from 2001 to 2002 and one large project beginning February 2006 and continuing through May, 2009.  The observations were made day and night.  Roughly 1000 galaxies have been observed between these programs.

The earlier observations were conducted during  commissioning of the GBT in the fall and winter of 2001/2 as a background 21-cm observing program during times when the telescope was not occupied with tests or calibration. Simple on-off spectral line measurements were made to acquire global HI profiles of galaxies at redshifts out to about 10,000 \kms.  Integration times were between 10 and 60 minutes and typical bandwidths were 5 or 10 MHz depending on the expected signal strength and line profile width. All observations used the FFT spectrometer which has 1024 channels for each of the two linear receiver polarizations. The system temperature was slightly under 20 K at high elevations and the 100-meter aperture efficiency was roughly 70\%.
Intensity calibration of the HI survey of galaxies during commissioning of the Green Bank Telescope (GBT) is tied to the NVSS flux density scale \citep{1998AJ....115.1693C}. About five dozen continuum sources with flux densities between 2.2 and 6.0 Jy were selected to avoid significant multi-source confusion with the nine-arcminute GBT beam. The continuum calibrators were observed with the same spectrometer and receiver configuration as was used to measure the HI line profiles in the survey, with the exception that the spectrometer bandwidth was always 40 MHz centered on 1403 MHz to span most of the range of frequencies observed with smaller bandwidths.

The calibrator observing sequence was 2 minutes off, 2 minutes on, and 2 minutes off source in spectral line mode. The first off position was 38 arcminutes toward lower right ascension than the source position, and the second off was the same distance toward greater right ascension. The hour angle track was, therefore, not exactly the same for the three observed positions for high declination objects, but this did not appear to degrade the spectral baselines significantly. The first task in the calibrator data reduction was to visually inspect the difference spectrum between the two off positions. The difference spectrum offset was typically less than about 60 mJy, as is expected from confusion noise with the GBT beam size, but a few offsets were as high as 300 mJy. These large offsets were possibly due to a moderately strong source in one of the off positions or, more likely, a bit of radiation from the Sun during the day. Since the observed source flux density was about 3 Jy, even the largest off-source baseline offset, after the two off spectra were averaged together, caused about 5\% error in the measure source intensity. More typically, this source of error amounted to less than 1\%. The statistics of the calibrator source measurements were not significantly improved by throwing out observations with larger off-position differences so all data were retained.

For the observations since 2006 use is made of
the single beam L band (1 to 2 GHz) receiver
and the spectral line spectrometer as the backend detector.
Data are taken with a 12.5 MHz bandwidth and 9-level sampling.
Total power observations are made with
a full calibration noise source switching cycle of 1 second.
The spectrometer records data every 30 seconds.
The spectral line is Doppler tracked in the barycentric velocity frame.
Data are taken using linear polarizations.
Redshifts from the Lyon Extragalactic Database (LEDA) or NASA Extragalactic Database (NED) were used
to center the window.

The basic GBT observation procedure was to take a pair of  on--off observations with 300s on and 300s off the target.
Galaxies within 3,000 \kms\ usually required 1 to 3 scan pairs, while galaxies from the PSCz sample reaching up to 8,000 \kms\
required 10 to 15 scan pairs. A preliminary guess on how much observing time a target would need was derived from
the 21 cm magnitude given in LEDA. In order to optimize the observing time, a target was observed with time split over several days. Data were reduced daily and evaluated
in order to add observing time as needed until the signal
reached the desired very high quality for a luminosity--linewidth distance measurement.

Individual observations are calibrated in Jy using the standard calibration procedure available at the GBT.  GBTIDL provides basic routines that can be used to
calibrate and average spectra when the data are taken in standard, predefined observing
modes. The calibration routines typically give a flux scale accurate to 10\%.
Well-known galaxies that can be used as HI calibrators were also observed several times per month in order to be able to retrieve
a more accurate flux calibration if needed in the future.

The calibrated data are then averaged, baseline subtracted using a polynomial fit usually of order 3, and smoothed with a simple Hanning filter.
The final spectrum is stored with 1.6 \kms\ resolution. It was usually binned at least once to 3.2 \kms\ resolution for the HI  linewidth measurement.

\subsection{Parkes}

Observations with Parkes Telescope make use of the 7--beam system in MX mode, with the target in the central beam and the 6 outer beams used to monitor the sky. Integration times are estimated based on fluxes obtained with HIPASS integrations of roughly 8 min per pointing, velocity resolution of 18 \kms, and r.m.s. sensitivity of 13 mJy per channel.  For linewidths adequate for our purposes we attempt to obtain spectra with peak signal to noise of at least 10 with spectral resolution of 2 \kms.  Galaxies at 3000--4000 \kms, which are the most common of our targets, typically require 60 min on source.  Profiles are evaluated at 30 min exposure and, if inadequate, the source is reobserved and profiles are summed.  Daytime observations are avoided to minimize degradation by the Sun. The Parkes multibeam data are reduced using the graphical user interfaces Livedata, Gridzilla, and MIRIAD.  The data is calibrated in Janskys using the procedure described by \citet{2001MNRAS.322..486B}.

At press, a first observing run during January--February 2009 has been completed culminating in observations of 58 galaxies.  Results from this run and the much larger archival sample of 1000 brightest sources from the HIPASS program \citep{2004AJ....128...16K} are analyzed and included in EDD.     

\section{Profile Linewidths}

The linewidth measure given in the {\it Pre-Digital HI} catalog of EDD and used by us since the early paper by  \citet{1977A&A....54..661T} is $W_{20}$, the linewidth at 20\% of peak intensity.  This is an appropriate moment to evaluate whether that parameter choice is optimal since we now do the analysis on digital data with a rigorous algorithm and apply the same procedure to all available material.  The thinking behind the original choice was that a linewidth at a very low level of intensity with respect to the maximum is desirable to minimize dependencies on vagaries in the distribution of flux within the profile. The opposing constraint is the need to be above the noise level.  With profiles deemed adequate, it was empirically determined that the 20\% of peak intensity level is sufficiently out of the noise.  Adequate profiles are characterized by peak signals at least seven times greater than the noise.

\begin{figure}[htb!]
\figurenum{3}
\centering
\includegraphics[scale=.7]{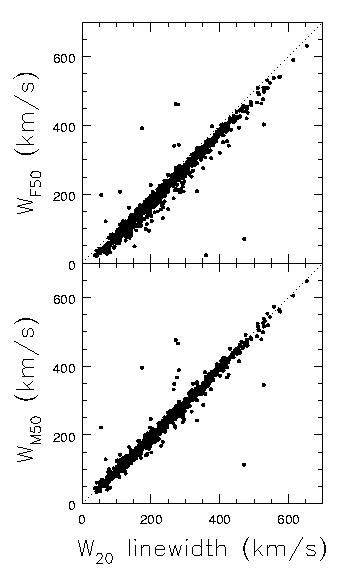}
\caption{Comparison between the $W_{20}$ linewidth parameter from the `Pre-Digital HI' catalog in EDD with two of the linewidth parameters given by SHGK, $W_{F50}$ and $W_{M50}$.  The comparison is based on 1110 cases with good $W_{20}$ measures.}
\label{w20_wx50}
\end{figure}

As we consider alternatives, we look to the study by \citet{2005ApJS..160..149S} [SHGK].  Their data are made available at the Cornell University Digital HI Archive website\footnote{http://arecibo.tc.cornell.edu/hiarchive}.  Those authors have given attention to a large body of high quality data from their own observations and from the archives.  They derived five separate linewidth parameters with automated algorithms.  These distinct linewidth measures can be compared with each other and, for most of the galaxies within 3,000 \kms, with the $W_{20}$ values from the {\it Pre-Digital HI} catalog in EDD.  Comparisons are shown in Figure~\ref{w20_wx50} for 1110 galaxies considered to have good $W_{20}$ measures.  In 3\% of cases the linewidths are discordant by more than 50~\kms.  These large differences are due to cataloging errors or gross errors due to noise.  Automated procedures are vulnerable to occasional gross errors -- as manifested by big  differences between the five SHGK parameters in a small fraction of cases.  In the following discussion, we clip instances with deviations greater than 50~\kms\ from the mean of $W_{20} - W_X$, where $X$ is one of the five linewidths given by SHGK.

\begin{figure}[htb!]
\figurenum{4}
\plotone{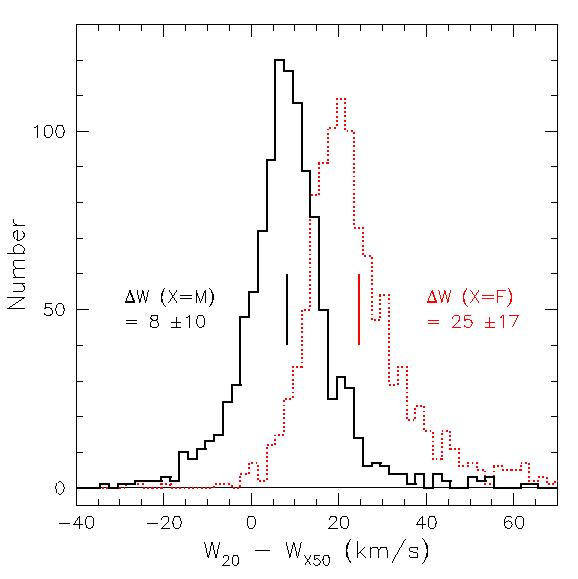}
\caption{Histograms of the differences in linewidth value $W_{20}$ given in the `Pre-Digital HI' catalog of EDD and the linewidth values $W_{X50}$ given by SHGK where $X=F$ (dotted histogram in red) is the parameter preferred by those authors, a measure at 50\% of peak flux, and $X=M$ (solid histogram in black) is a measure at 50\% of the mean flux.  The histogram of $W_{20} - W_{F50}$ values is skewed with a tail to positive differences and a relatively large dispersion.  The histogram of $W_{20} - W_{M50}$ values is symmetric and has lower dispersion.}
\label{hist_w20_wx50}
\end{figure}

It turns out that we find a clear preference for one of the SHGK linewidth parameters.  It is {\bf not} the parameter advocated as optimal by SHGK.  Those authors prefer the parameter $W_{F50}$,
the width at 50\% of the peak minus rms flux with left and right edges evaluated independently
with polynomial fits to the rising portions of the profile.  In Table~1 we collect comparisons between our old $W_{20}$ values and the five SHGK values (plus their parameter $W_C$ which is  $W_{F50}$ with redshift, instrumental, and smoothing corrections).  Of course, zero-point offsets are expected.  The figure-of-merit  is the r.m.s. dispersion.  The SHGK parameter $W_{M50}$ gives a significantly better correlation with our $W_{20}$.  Conveniently, it also gives rather close agreement in zero-point.
Figure~\ref{hist_w20_wx50} gives a detailed comparison between $W_{20}$ and the two SHGK linewidth parameters of greatest interest.  In the case of the parameter $W_{F50}$ the mean difference with respect to $W_{20}$ is displaced from the peak of the distribution (the distribution is skewed)
and (after clipping values more deviant than $\pm 50$~\kms\ from the mean) the dispersion is a rather substantial 17~\kms.  In the case of the parameter $W_{M50}$ the histogram is symmetric and (after clipping values more deviant than $\pm 50$~\kms\ from the new mean) the dispersion is a reasonable 10~\kms.

\begin{deluxetable}{lcc}
\tablenum{1}
\tablecaption{Differences between Pre Digital and SHGK Linewidth Measures}
\label{tbl:delw}
\tablewidth{0in}
\tablehead{\colhead{$W_{20} - W_X$} & \colhead{Mean Diff.} & \colhead{R.M.S.}}
\startdata
$X = F50$ & 25 \kms & 17 \kms  \\
$X = M50$ & ~8 \kms & 10 \kms \\
$X = P50 $ & 26 \kms & 21 \kms \\
$X = P20 $ & -7 \kms & 22 \kms \\
$X = 2P50 $ & 14 \kms & 17 \kms \\
$X = C $ & 32 \kms & 17 \kms \\
\hline
$X = m50$ & 15 \kms & 11 \kms \\
\enddata
\end{deluxetable}

The parameter $W_{M50}$ is the linewidth at 50\% of the mean flux level within the HI signal.  This construct has nice features.  Using the mean flux level rather than the peaks serves to disengage the linewidth measure from details of the gas distribution.  It gives a more natural transform over to single peak cases.  The measurement is at a low level compared with the peak, statistically only slightly above the level that gives $W_{20}$ ( hence the similar zero-point).  The main operational challenge is to define the window containing the signal.
SHGK reserve most of their discussion for their linewidth parameter $W_{F50}$ and do not give details of the derivation of $W_{M50}$.  We have developed an algorithm which might differ in minor details.  An empirical test of that possibility will come from inter comparisons of our separate results with the same data, reported in the next section.  To distinguish our parameter from that developed by SHGK, we refer to our measurement as $W_{m50}$, that is, with a lower case `$m$'.

The most sensitive detail with the derivation of $W_{m50}$ is the specification of the wavelength window for the summation of flux.  The total flux detected within the single beam pointing is relatively well defined.  However the mean flux per channel can be significantly less well defined if channels in the wings of the profile are included or not because of noise.

Our specific recipe is to first determine the integrated flux within a window that is tight, yet sufficiently wide that the profile has reached the baseline level (call this the 100\% window).  Then the wavelengths are determined that exclude 5\% of the integrated flux on each of the two wings (these enclose the 90\% window).  The mean flux per channel is then taken to be the sum of the flux within this 90\% window divided by the number of spectral channels.  Both the numerator and denominator in this calculation require interpolations since the wavelengths defining the window are not restricted by the discreteness of the spectral channels.  

The linewidth is defined at the level of 50\% of the mean flux per channel determined in the manner that has just been described.  The intersection points at the two edges are defined by interpolations between observed flux--velocity points on the rising parts of the profile.  SHGK use a more elaborate fitting scheme to define edges but it is not clear that the effort results in greater precision.   In clean cases the spectra rise abruptly and in ambiguous cases the uncertainty is not necessarily ameliorated by a particular fitting recipe.

The assignment of errors is a particularly challenging problem.  Profiles can be messy in so many ways that we despaired of finding an algorithm that gives sensible results in all cases.  Our overwhelming interest is the use of profile widths as a parameter in the measurement of distances.  We consider that there is a threshold of acceptability; a profile may be of sufficient quality to be used in the determination of a distance, or it may not be.  We link our error estimate to this threshold.  Specifically, an adequate profile is assigned an error of less than or equal to 20~\kms.  Inadequate profiles are identified by errors greater than 20~\kms.  It has been found \citep{2000ApJ...533..744T} that with linewidth errors constrained to this limit in quality the measure of linewidths is not a dominant source of errors in the determination of distances.

The detailed error assignment is arrived at in two steps.  The first step is automatically generated based on the signal (the mean flux per channel within the 90\% window), $S$, to r.m.s. noise, $N$.  Errors were evaluated from a training set.  Errors of 8~\kms\ are assigned in the best cases, whenever the mean flux per channel is greater than 17 times noise.  Errors degrade to 20~\kms\ by a mean flux per channel to noise of 2, and continue to increase as signal degrades further.  Specifically, the error $e_W$ is assigned based on signal-to-noise, $S/N$, following:

$e_W = 8$ \kms\ ~~~~~~~~~~~~~~~~~~~~~~~~~~~~~~~~ if  $ S/N > 17$

$e_W = 21.6 - 0.8~S/N$ \kms\  ~~~~~~~~~~~~~ if $ 2 < S/N < 17$

$e_W = 70 - 25~S/N$ \kms\ ~~~~~~~~~~~~~~~~  if $ S/N < 2$
 
\noindent
Errors are not allowed to be less than the spectral resolution after smoothing.  Our error estimates are conservative; roughly a factor 2 larger than values found in the literature associated with the same observations.  Cross-correlations between data sets to be discussed later will demonstrate that our error estimates must on average be $\sim 50\%$ greater than a $1 \sigma$ value.

The second step involves a manual inspection, given to every profile.  The fit and error estimate proposed by the computer algorithm is displayed on a monitor.  The operator can accept or modify.  It may be necessary to excise interference, or reposition the edges of the 100\% window (the most common occurrence), or smooth.  The computer redisplays.  The final decision by the operator is whether the error is appropriate, which is a binary decision to the question: is the profile adequate or inadequate for the purpose of measuring a distance?  The error only needs to be changed one way or the other across the 20~\kms\ acceptance threshold if the carbon-based decision contradicts the silicon-based assignment.

\subsection{Good, Bad, and Ugly}

\begin{figure}[htb!]
\figurenum{5}
\plotone{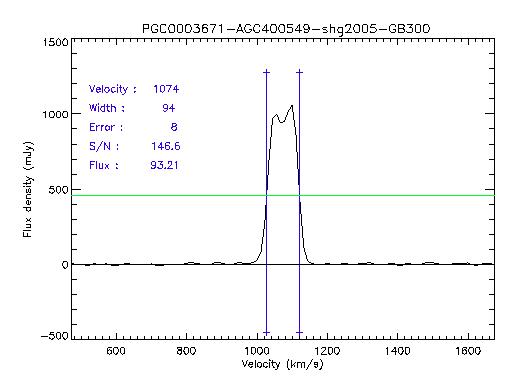}
\plotone{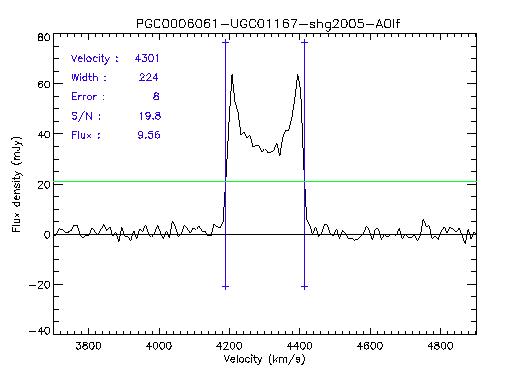}
\plotone{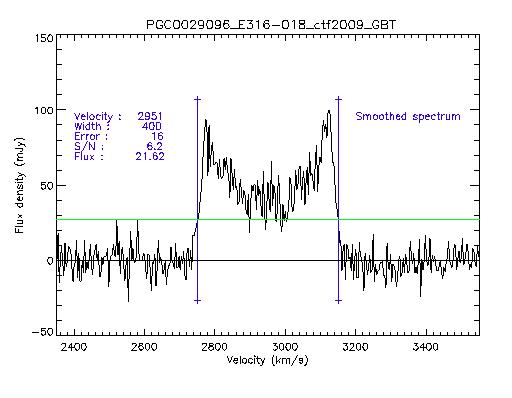}
\plotone{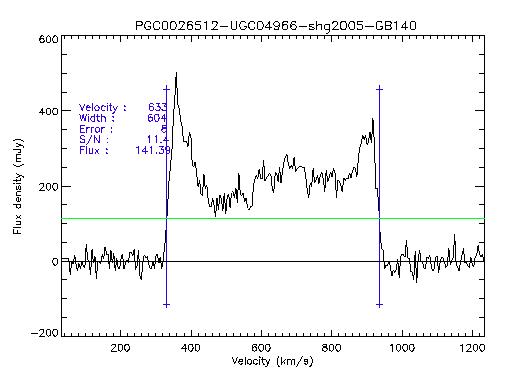}
\caption{PGC 3671 (NGC 337A),  PGC 6061 (UGC 1167), PGC 29096 (ESO 316-018), and PGC 26512 (NGC 2841), profiles with increasing linewidths, observed with the NRAO $300^{\prime}$, Arecibo, Byrd Green Bank, and NRAO $140^{\prime}$ telescopes respectively.}
\label{4good}
\end{figure}

\begin{deluxetable}{llll}
\tablenum{2}
\tablecaption{Telescopes Contributing to the Database}
\label{tbl:tel}
\tablewidth{0in}
\tablehead{\colhead{Telescope} & \colhead{Aperture} & \colhead{Beam} & \colhead{Acronym}}
\startdata
Arecibo     & 305m       & ~$3^{\prime}$       & AOG-AOlf-ALFA \\
Nan\c cay & 200x40m & ~$4^{\prime}$ x $22^{\prime}$  & nan \\
GBT           & 100m        & ~$9^{\prime}$    & GBT \\
Effelsberg & 100m        & ~$9^{\prime}$       & Effs-Eff \\
GB300       & ~91m        & $10^{\prime}$      & GB300 \\
Parkes       & ~64m        & $14^{\prime}$     & PAKS \\
GB140       & ~43m        & $21^{\prime}$      & GB140 \\
\enddata
\end{deluxetable}

\begin{deluxetable}{ll}
\tablenum{3}
\tablecaption{All Digital HI Catalog Sources}
\label{tbl:sources}
\tablewidth{0in}
\tablehead{\colhead{Code} & \colhead{Literature Source}}
\startdata
ksk2004 & Koribalski, Staveley-Smith, Kilborn, et al. 2004 \\
shg2005 & Springob, Haynes, Giovanelli, Kent 2005 \\
hkk2005 & Huchtmeier, Karachentsev, Karachentseva, et al. 2005 \\
tmc2006 & Theureau, Martin, Cognard, et al. 2006 \\
ghk2007 & Giovanelli, Haynes, Kent, et al. 2007 \\
sgh2008 & Saintonge, Giovanelli, Haynes, et al. 2008 \\
kgh2008 & Kent, Giovanelli, Haynes, et al. 2008 \\
ctf2009 & Courtois, Tully, Fisher, et al. 2009 (this paper)\\
\enddata
\end{deluxetable}

Examples of good profiles are seen in Figure~\ref{4good}.  They are characterized by high flux `horns' at the extrema of the profiles, which arise from emission from the flat portion of rotation curves near maximum velocity.  The profiles rise rapidly, leaving little uncertainty in the measurement of the widths.
There is apparent splitting into two peaks even with the narrowest profile chosen as an example in this figure although typically the distinctness of two peaks is lost in narrow profiles. In this figure and others to follow, information regarding the source of the profile is encoded in the header above the profile.  One is given the PGC number, a common name, the archival or new observation source in a code given in Table 3 and the telescope as identified in Table 2.  

\begin{figure}[htb!]
\figurenum{6}
\plotone{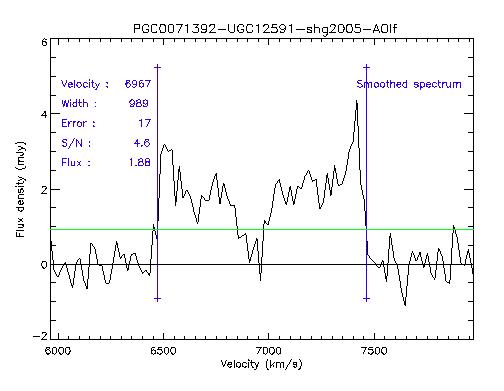}
\plotone{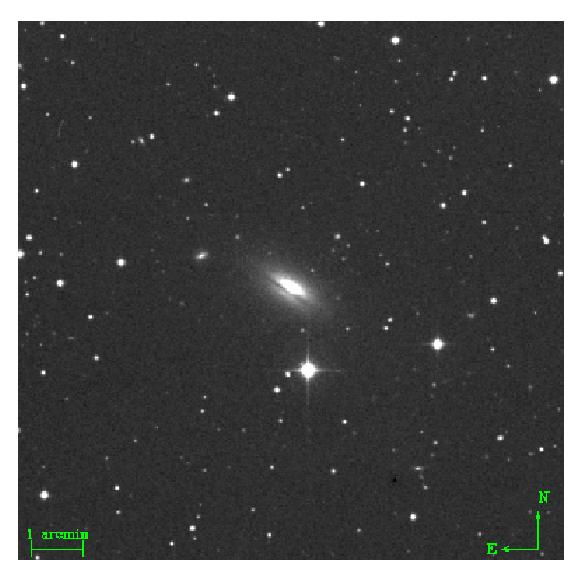}
\caption{PGC 71392 (UGC 12591), the galaxy with the largest linewidth with $W_{m50} = 989$~\kms\ at $V_h = 6967$~\kms.  The velocity scale for this exceptionally wide profile is expanded. Observed with Arecibo Telescope.}
\label{p0071392}
\end{figure}

The catalog {\it All Digital HI} now contains information on approximately 13,000 galaxies.  Of reliably measured profiles for apparently single targets, that for PGC 71392 (UGC 12591) is by far the widest, with $W_{m50} = 989$~\kms.  The line profile and an image of the galaxy (typed SO/Sa) are shown in Figure~\ref{p0071392}.  \citet{1986ApJ...301L...7G} have drawn attention to this unusual galaxy.  That reference notes that there is probably absorption from a central continuum source affecting the spectrum.

\begin{figure}[htb!]
\figurenum{7}
\plotone{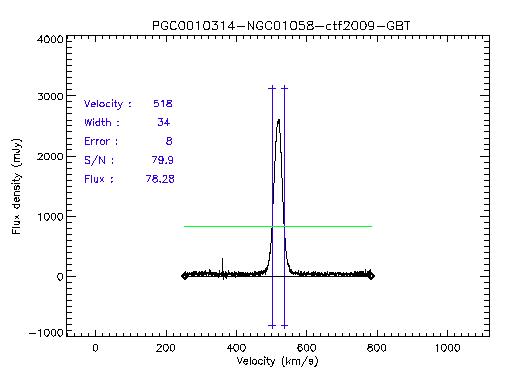}
\plotone{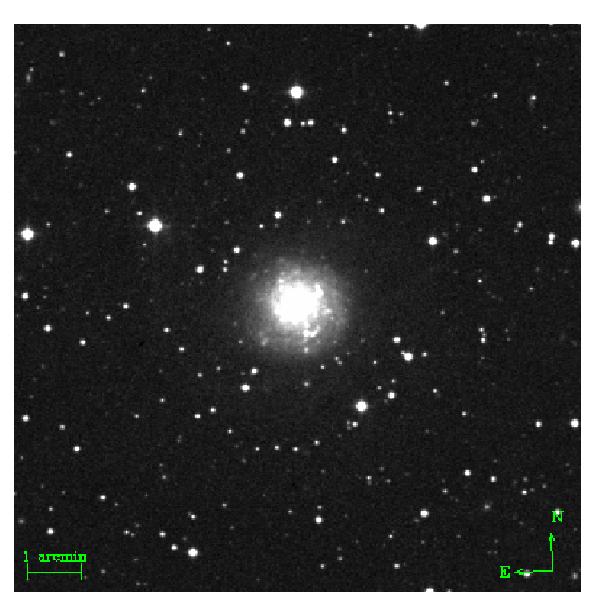}
\caption{PGC 10314 (NGC 1058), the galaxy with the narrowest linewidth with $W_{m50}^c = 33$~\kms. Observed with GBT.}
\label{p0010314}
\end{figure}

At the other extreme, PGC 10314 (NGC 1058) is the most anorexic of galaxies in the current database.  The profile is shown in Figure~\ref{p0010314}.  This galaxy has been identified by \citet{1975A&A....44..147L} and \citet{1984A&A...134..258V} as an example of a galaxy seen almost face on.

There are lots of bad spectra among the 15,000 profiles in {\it All Digital HI}.  The two main reasons for bad spectra are poor signal-to-noise and confusion from multiple sources in the radio beam.  It is easy to find examples of poor spectra; select cases with error assignments $e_W > 20$~\kms.  We adopt the convention of assigning $e_W = 100$~\kms\ in cases of confusion, and $e_W = 500$ \kms\ in cases of null detections, although we have not been consistent.  Our fundamental convention is, if the profile is inadequate to the task of measuring a distance, then an error greater than 20~\kms\ is assigned.  The exact value of an error assignment greater than 20~\kms\ has little rigor. 

\begin{figure}[htb!]
\figurenum{8}
\plotone{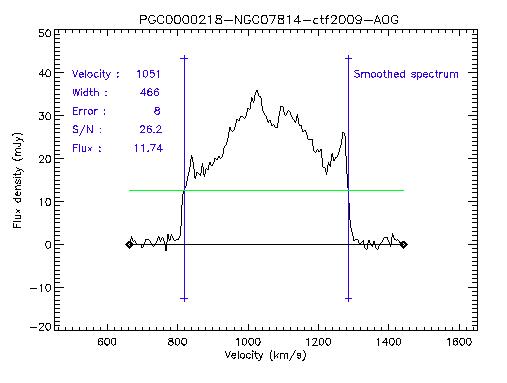}
\plotone{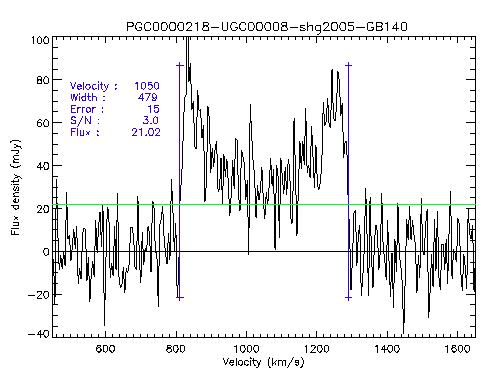}
\plotone{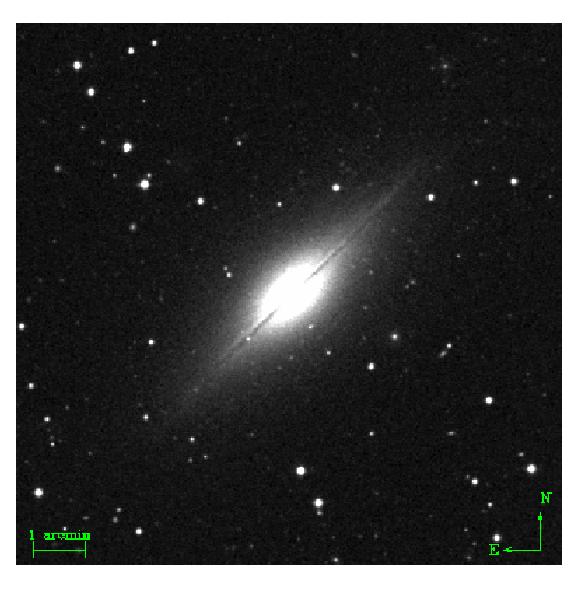}
\caption{PGC 218 (NGC 7814).  Top profile: The $3^{\prime}$ half-power beam of Arecibo Telescope is smaller than the target galaxy causing flux to be lost and affecting the profile shape.  Bottom profile: All the flux emitted by the galaxy is captured within the beam of the Green Bank 140-foot Telescope.}
\label{p0000218}
\end{figure}

While it is not worthwhile to dwell on the bad, it is instructive to consider a few examples of the ugly.  A cautionary example is illustrated in Figure~\ref{p0000218}.  The two profiles were obtained with Arecibo Telescope and the Green Bank 140-foot Telescope with, respectively, half-power beams of $3^{\prime}$ and $21^{\prime}$.   The galaxy NGC~7814 has a diameter at the $B$ band isophot of 25~mag~as$^{-2}$ of $5.5^{\prime}$, larger than the Arecibo beam but much smaller than the 140-foot beam.  Flux is lost in a single beam pointing with Arecibo Telescope but not with a pointing involving the smaller telescope.  The lost flux from the extremities of the galaxy with the Arecibo observation cause a pronounced reduction in the `horns' and affects the measurement of the profile width.

\begin{figure}[htb!]
\figurenum{9}
\plotone{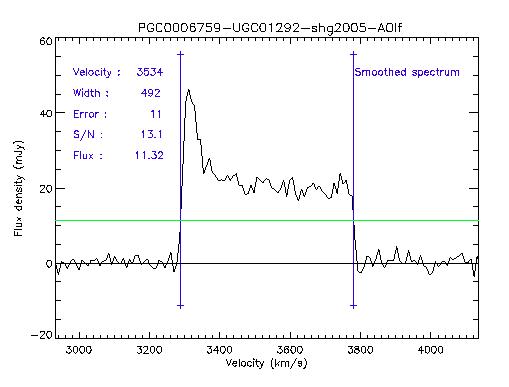}
\plotone{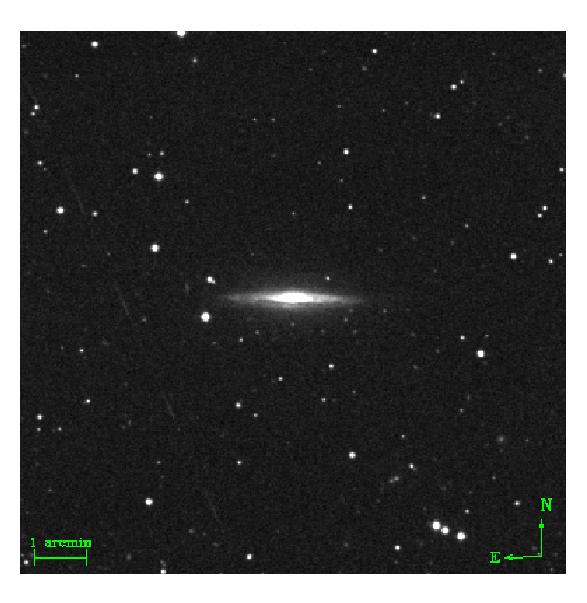}
\caption{PGC 6759 (NGC 684).  One HI peak is much more pronounced than the other.  This profile was obtained with Arecibo Telescope.}
\label{p0006759}
\end{figure}

In Figure~\ref{p0006759} one sees what seems to be a normal edge-on spiral but one horn is very pronounced and the other is almost unseen.  A profile like this creates a problem if the measurement of the width is referenced to the peak flux.  It creates somewhat less problem with our derivation based on the mean flux.

\begin{figure}[htb!]
\figurenum{10}
\plotone{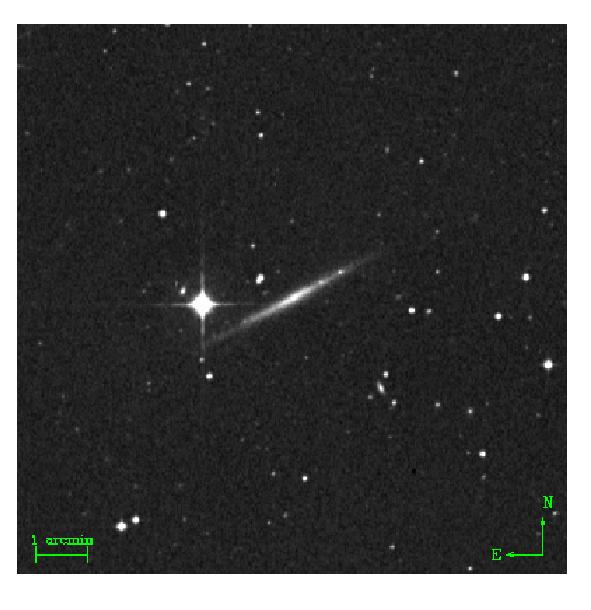}
\includegraphics[scale=0.5]{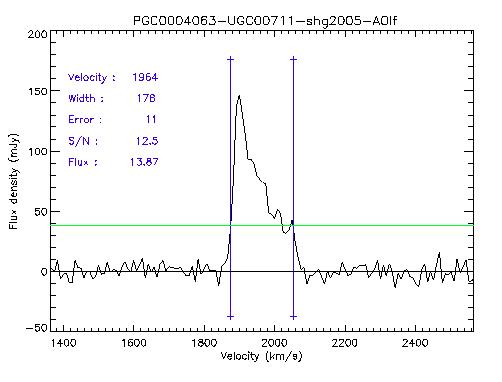}
\includegraphics[scale=0.5]{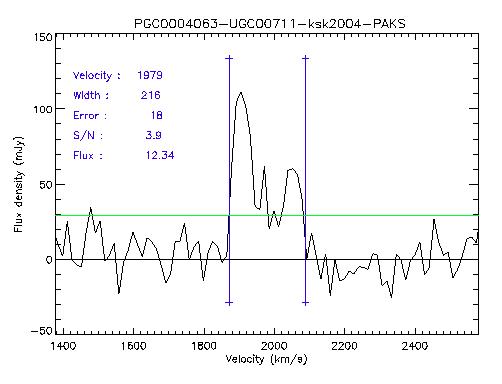}
\caption{PGC 4063 (UGC 711). Profile on top obtained with Arecibo Telescope is extremely asymmetric and is suspected to be biased because of beam attenuation. Profile on the bottom was obtained with the 5 times larger beam of the Parkes Telescope.  The asymmetry of the profile is not just a resolution effect.}
\label{p0004063}
\end{figure}

The situation is even more extreme with the galaxy shown in Figure~\ref{p0004063}. In this case the distribution of neutral hydrogen in the Arecibo profile is so asymmetric that the long wavelength edge of the profile is poorly defined. Given the size of the galaxy, the Arecibo observation may be missing flux.  The profile obtained with Parkes Telescope is asymmetric in the same sense though less extreme. The $14^{\prime}$ primary beam of the Parkes Telescope is considerably bigger than the source so the asymmetry is undoubtedly a real feature of this galaxy.  In this case, the Parkes profile is preferred over the Arecibo profile. 

\begin{figure}[htb!]
\figurenum{11}
\plotone{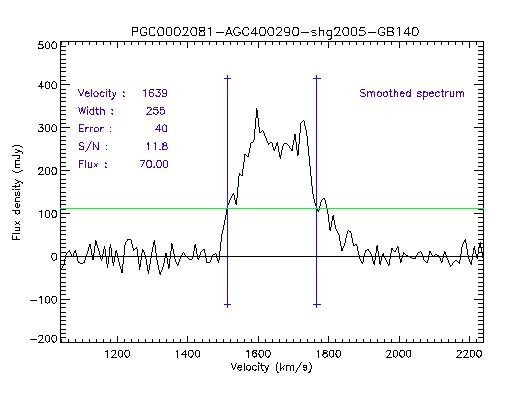}
\plotone{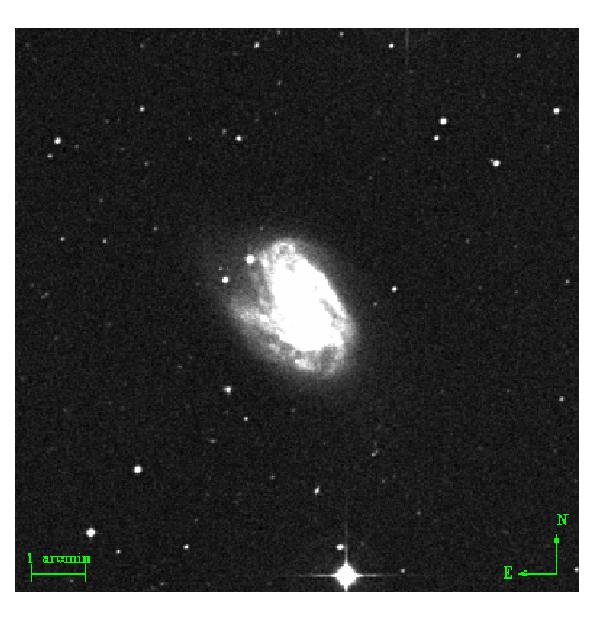}
\caption{PGC 2081 (NGC 157).  Wing on right edge of profile.  The galaxy is somewhat distorted. The observation was made with the Green Bank 140$^{\prime}$ telescope.}
\label{p0002081}
\end{figure}

In the case shown in Figure~\ref{p0002081} the profile has gone beyond ugly to bad.  There is a substantial wing on the long wavelength side.  The linewidth is acutely sensitive to the choice of level of measurement.  The galaxy looks distorted.  A galaxy interaction is suspected.

\begin{figure}[htb!]
\figurenum{12}
\plotone{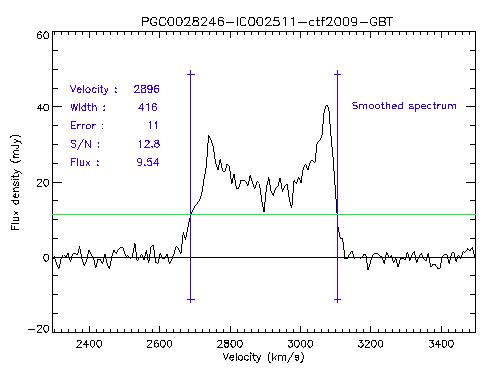}
\plotone{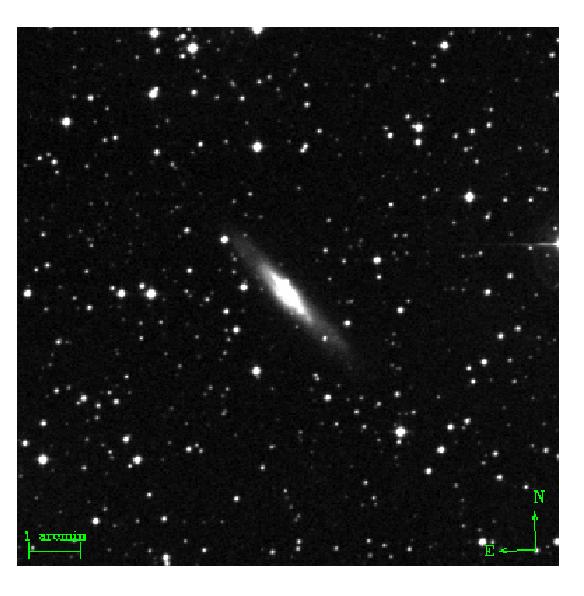}
\caption{PGC 28246 (IC 2511). Messy wing on left side of profile obtained with GBT.}
\label{p0028246}
\end{figure}

With IC~2511 seen in Figure~\ref{p0028246} the profile is merely ugly.  There is a wing on the short wavelength side.  The linewidth is sensitive to the choice of measurement flux level, but not sufficiently to cause us to reject the profile by assigning an error greater than 20~\kms.  There is no evidence of an abnormality in the image of the galaxy.  It can be appreciated that there is a continuum of situations between those shown in Figs.~\ref{p0002081} and \ref{p0028246}, commonly aggravated by much worse confusion from noise.  In the final analysis, profiles have been accepted or rejected (assigned errors less or greater than 20~\kms) on the basis of visual inspection.   

\begin{figure}[htb!]
\figurenum{13}
\plotone{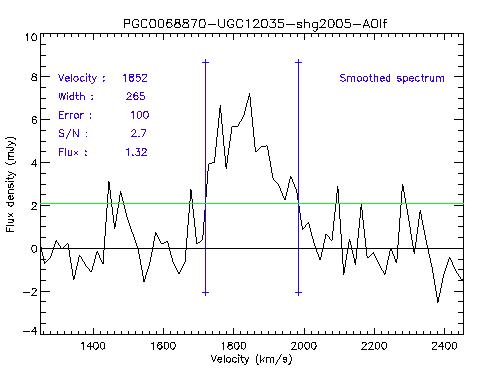}
\plotone{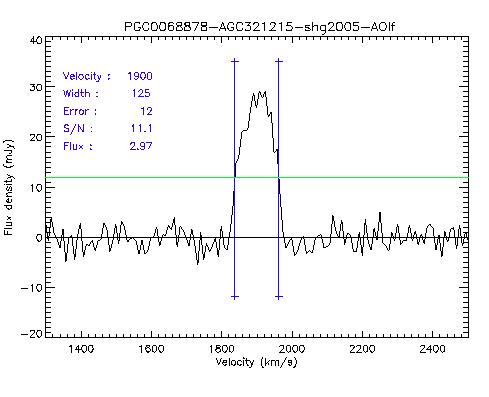}
\plotone{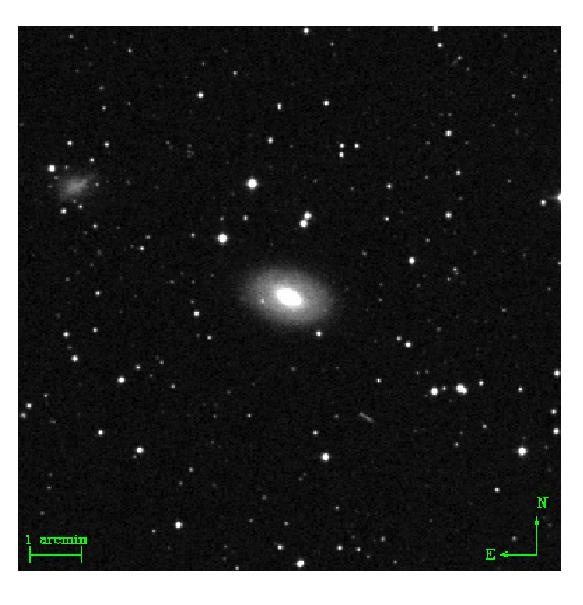}
\caption{PGC 68870 (NGC 7280) and PGC 68878 (UGCA 429) separated by $4.7^{\prime}$.  Arecibo observations of each resolves the separate sources.  UGCA 429 at upper left has the stronger, unambiguous signal.  NGC 7280 in the center of the image is clearly detected but flux in the right wing must be associated with UGCA 429.}
\label{p0068870}
\end{figure}

An example of contamination from multiple sources in the radio beam is provided by Figure~\ref{p0068870}.  Both the galaxy at the center of the image and the fainter object $4.7^{\prime}$ northeast have been observed with separate pointings with the Arecibo Telescope.  The half-power beam diameter with that telescope is 3 arcmin.  One of the sources is cleanly detected with a peak flux four times greater than the other.  Lo and behold, it is the smaller fainter galaxy to the northeast.  The visibly dominant galaxy is detected in HI but the profile is messy and probably contaminated by flux from the companion.  We assign an error of 100~\kms\ (confused) to the linewidth of the brighter galaxy. 

\begin{figure}[htb!]
\figurenum{14}
\plotone{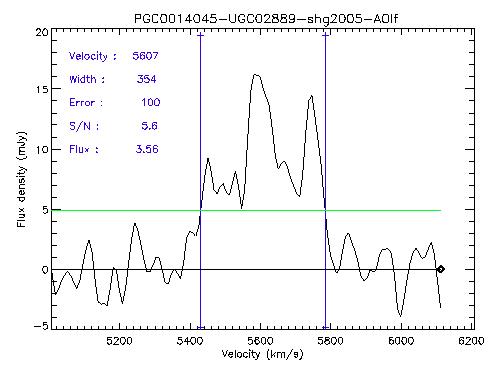}
\plotone{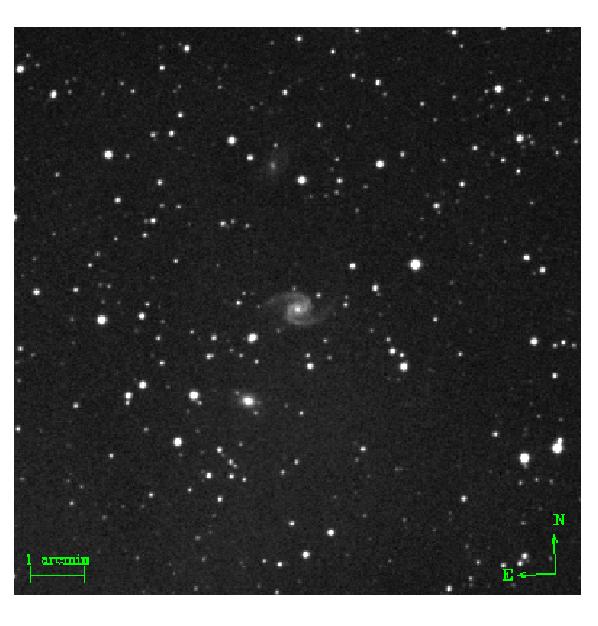}
\caption{PGC 14045 (UGC 2889) observed with Arecibo Telescope.  Anomalous profile with central peak.  Small galaxy with unknown velocity $2.0^{\prime}$ SW might contaminate.}
\label{p0014045}
\end{figure}

Another all-too-common situation is illustrated in Figure~\ref{p0014045}.  The profile is anomalous, with a pronounced peak and shoulders.   In the image, a second galaxy of unknown velocity is seen that would lie near the half-power level of the Arecibo Telescope beam.  A third galaxy, fainter and more distant, could conceivably contribute to the confusion.  

These examples give a reminder of the advantages of observations with different facilities.  The relevant telescopes and the single beam half power field diameters are listed in Table~\ref{tbl:tel}.  Nan\c cay Telescope has an unusual beam shape that provides good resolution east-west but poor resolution north-south.  Arecibo Telescope provides the best resolution but attention must be given to possible loss of flux with large targets.

The examples given attention in this section provide the warning that there are lots of unacceptable line profiles for our purposes.  Even among those identified as acceptable by the error estimate, comparisons when alternative observations are available reveal that 2--3\% are bad.  Still, among 15,000 profiles there are an abundance of good data.  In the next section there is an evaluation of how good is good.

\section{Evaluation of the New $W_{m50}$ Parameter}

\begin{figure}[htb!]
\figurenum{15}
\includegraphics[scale=0.33]{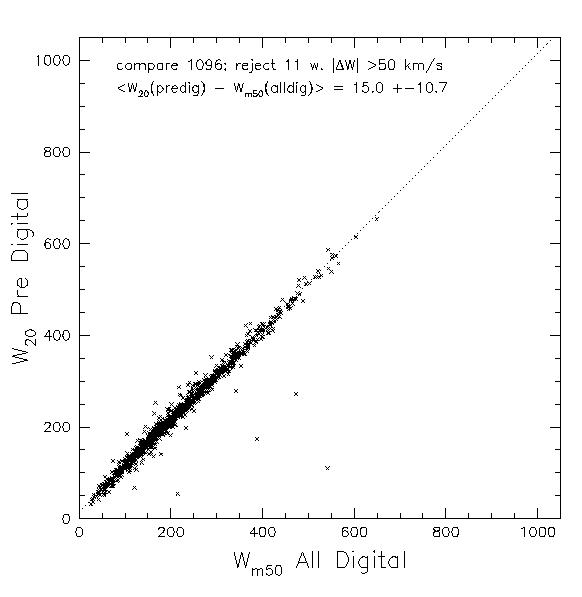}
\includegraphics[scale=0.33]{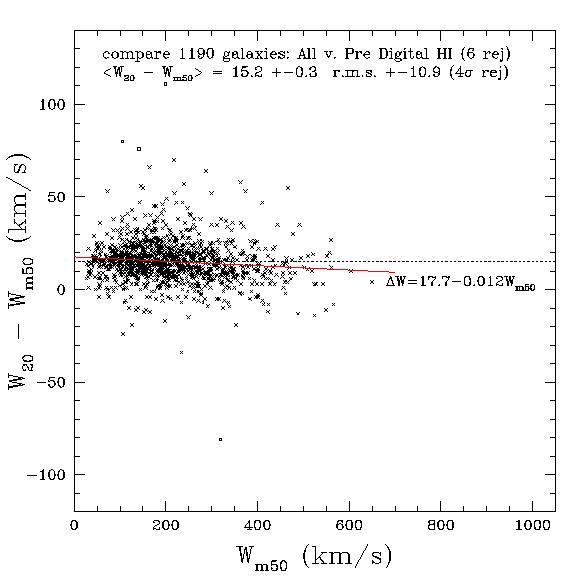}
\includegraphics[scale=0.4]{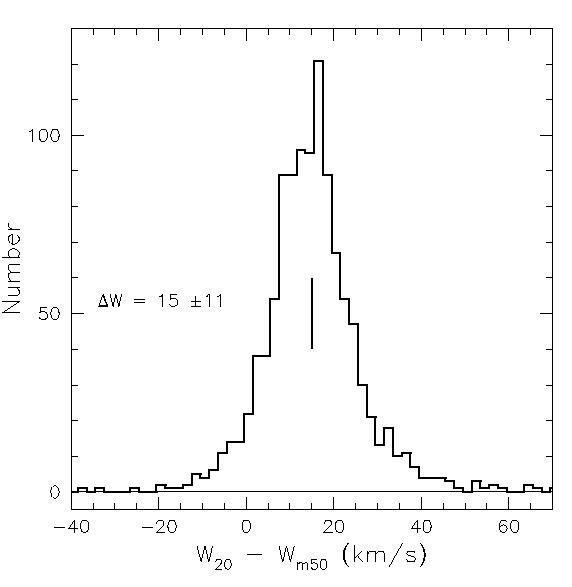}
\caption{Top:  Comparison between the linewidth at 50\% of mean flux in the catalog {\it All Digital HI} and the linewidth at 20\% of peak intensity in the catalog {\it Pre Digital HI}.   
Middle: Linewidth difference as a function of $W_{m50}$ with a least squares fit superposed.
Bottom: Histogram of differences $W_{20} - W_{m50}$.}
\label{w20-wm50}
\end{figure}

\begin{figure}[htb!]
\figurenum{16}
\includegraphics[scale=0.33]{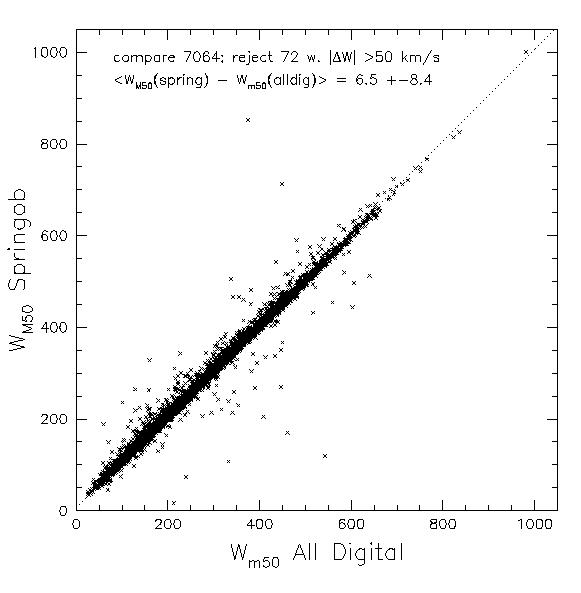}
\includegraphics[scale=0.33]{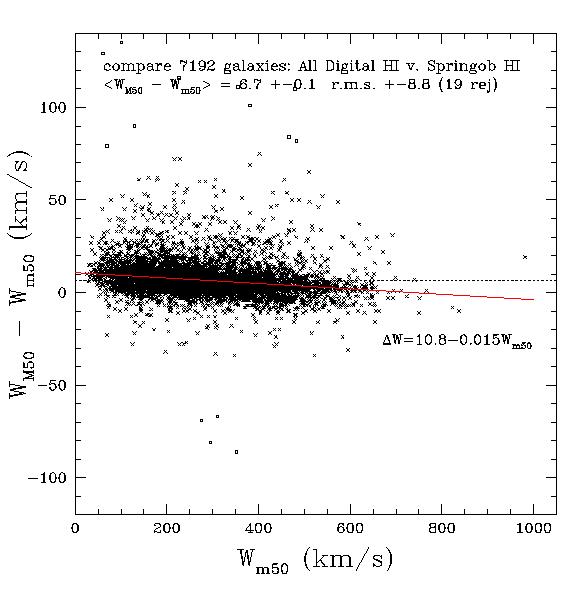}
\includegraphics[scale=0.39]{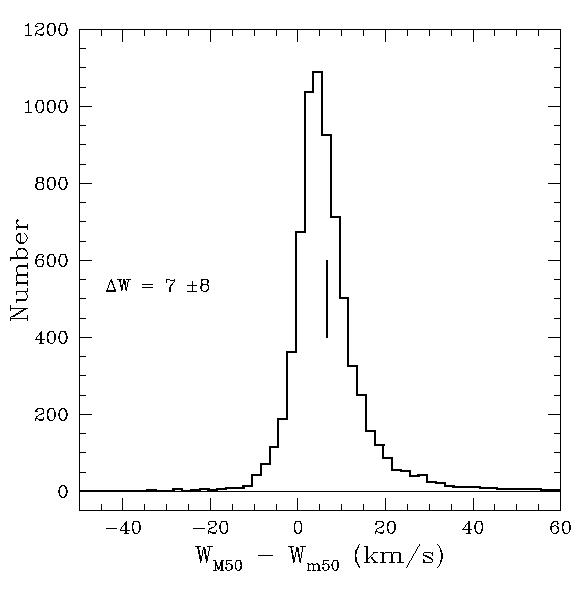}
\caption{Top: Comparison between two alternative estimators of the linewidth at 50\% of mean flux, $W_{m50}$ from the {\it All Digital HI} catalog and $W_{M50}$ from the {\it Springob/Cornell HI} catalog.  
Middle: Linewidth differential plot with least squares fit as a function of $W_{m50}$.
Bottom: Histogram of differences $W_{M50} - W_{m50}$.  The distribution is skewed, with a median difference of 5~ \kms.}
\label{wM50-wm50}
\end{figure}

Comparisons between alternative linewidth parameters are illustrated in Figures~\ref{w20-wm50} and \ref{wM50-wm50}.   In each case, measures with uncertainties greater than 20~\kms\ are rejected so the comparisons are between data that are supposed to be good.  In Fig.~\ref{w20-wm50},  the comparison is between our $W_{m50}$ parameter reported in the {\it All Digital HI} catalog and the $W_{20}$ parameter in {\it Pre Digital HI}.  In the top left panel, the $W_{m50}$ values are derived exclusively from data extracted from the Cornell HI archive.  Of 11 cases with linewidth measures that deviate by more than 50~\kms\ from the mean in a sample of 1107 galaxies, five can be traced to confused profiles caused by near neighbors.   The mean difference of $<W_{20} - W_{m50}> = 15$~\kms\ with the 1096 remaining galaxies  is expected since $W_{20}$ is measured at a fainter level.   The r.m.s. scatter after elimination of the 11 most deviant cases is a reasonable 11~\kms.    In the top right panel, there is a slight augmentation of the sample through the inclusion of new data acquired by the authors.  By plotting the difference in linewidth measures on the ordinate there is sufficient scale resolution to detect a weak dependency in the difference in linewidths with the amplitude of rotation: $W_{20}-W_{m50}=17.7-0.012W_{m50}$.  In the lower panel, it is seen that the histogram of the differences $W_{20} - W_{m50}$ is symmetric about the mean.   The details of this comparison are reported at the bottom of Table~1.

The comparison in Fig.~\ref{wM50-wm50} involves two separate algorithms to determine the linewidth at 50\% of mean flux.   There is the parameter given by SHGK referred to as $W_{M50}$ and the variant determined by our procedure reported in {\it All Digital HI} called $W_{m50}$.  As in the previous figure, the data displayed in the top left  panel draws exclusively from the Cornell HI archive while the data used to generate the right panel is slightly augmented by new observations.  The results are substantially the same. 
There is an offset of 6.6~\kms\ between the SHGK parameter $W_{M50}$ and our $W_{m50}$.  The SHGK parameter is evaluated at a slightly lower flux, presumably because they evaluate the mean flux over a wider wavelength window than our 90\% window.  There is a small but significant dependence of the offset on rotation rate: $W_{M50}-W_{m50}=10.8-0.015W_{m50}$.
Again, 1\% of cases are deviant by greater than 50~\kms, usually because of confusion caused by a companion.  Those aside, the r.m.s. difference from the mean is a satisfactory 9~\kms.  However as seen in the lower panel, the distribution is slightly skewed, with a tail to positive differences.  

Besides the Cornell HI archive, another important source of HI profile information is the LEDA database \citep{2003A&A...412...45P} and the related HI archive\footnote{http://klun.obs-nancay.fr} associated with the `Kinematics of the Local Universe' (KLUN) project  \citep{2006ASPC..351..429T}.  Derivative parameters and a discussion of results are presented by \citet{2007A&A...465...71T}.    Access to the KLUN tabular material and profiles is provided in EDD through the {\it HI Nancay} catalog.

A comparison with the KLUN results is not straight forward because of mixed use of the optical and radio conventions for transforming doppler shifts to velocities.  In the optical convention, which we use, one considers the shift in wavelength with respect to the rest value, $V_{opt} = c (\lambda - \lambda_0) / \lambda_0$,  while in the radio convention one considers the shift in frequency, $V_{rad} = c (\nu_0 - \nu) / \nu_0$.  Profiles displayed in the Nan\c cay database are presented in the radio convention although it is to be noted that the same profiles made available through NED, the NASA/IPAC Extragalactic Database\footnote{nedwww.ipac.caltech.edu/forms/SearchSpectra.html}, have been converted to the optical convention.  The tabular information presented by  \citet{2007A&A...465...71T} is mixed.  Systemic velocities have been transformed to the optical convention but linewidths appear to have remained in the radio convention.  For a galaxy at 7000~\kms\ with a linewidth $\sim 400$~\kms\ the linewidth in the optical convention is $\sim 20$~\kms\ wider than in the radio so the issue is significant.

\begin{figure}[htb!]
\figurenum{17}
\plotone{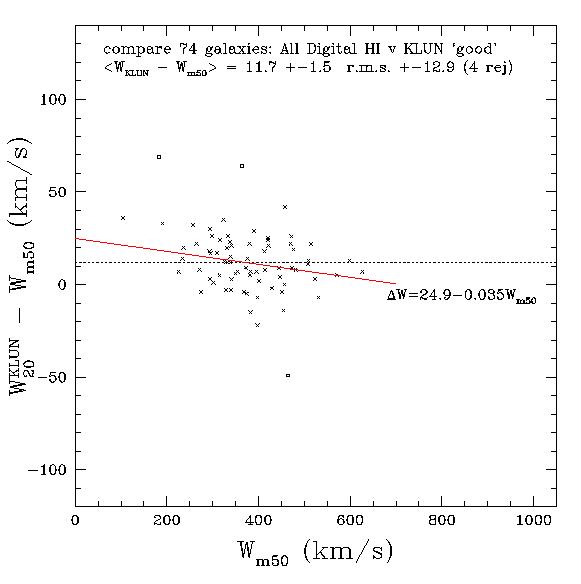}
\caption{Difference between the Nan\c cay/KLUN linewidth at 20\% of peak intensity and the $W_{m50}$  linewidth parameter in the {\it All Digital HI} catalog plotted against the $W_{m50}$ parameter.  A least squares fit is superimposed.  The slope has only a $2.2 \sigma$ significance.
}
\label{wnan-wm50}
\end{figure}

We have made comparisons between KLUN linewidths at 20\% of peak intensity, adjusted to the optical convention, and $W_{m50}$ linewidth values drawn from the {\it All Digital HI} catalog.  The comparison accepts only cases from the {\it All Digital HI} catalog with linewidth errors $\le 20$~\kms\ and cases from the {\it HI Nancay} catalog with a linewidth quality index assigned by us of 1-3.  See Figure~\ref{wnan-wm50}.  After rejection of 4 extreme outliers, the scatter is 13~\kms.  There is a hint of a correlation with $W_{m50}$ as found in Figs. \ref{w20-wm50} and \ref{wM50-wm50}.  The comparison sample is restricted because for the moment only a small fraction of galaxies in the {\it HI Nancay} catalog have been assigned a quality index by us.  If galaxies without a quality index are accepted then the comparison can be based on an order of magnitude larger sample of 992 galaxies.  However the scatter is then much worse.  Even after clipping 48 cases with deviations greater than 80 \kms\ from the mean the scatter is a poor 22~\kms.  There will be a discussion in the next section of the results of our own analysis of profiles extracted from the Nan\c cay database.  It will be seen that when the data are treated in a uniform way there is good agreement between all sources, whether drawn from the Nan\c cay or Cornell databases or derived from our new observations with GBT and Arecibo.  

The weak but significant tilt in the difference plots as a function of rotation rate seen in Figs.~\ref{w20-wm50} -- \ref{wnan-wm50} is a consequence of a different linewidth definition.  Our linewidths are measured at relatively higher flux levels for small galaxies and approach the levels of the other measures for large galaxies.  A result could be a slightly flatter slope for the luminosity--linewidth correlation.  These small differences serve to emphasize that the linewidth measures are empirical constructs.  We are reminded that the details of the construct may not be important but consistency is required if biases are to be avoided.

\begin{figure}[htb!]
\figurenum{18}
\plotone{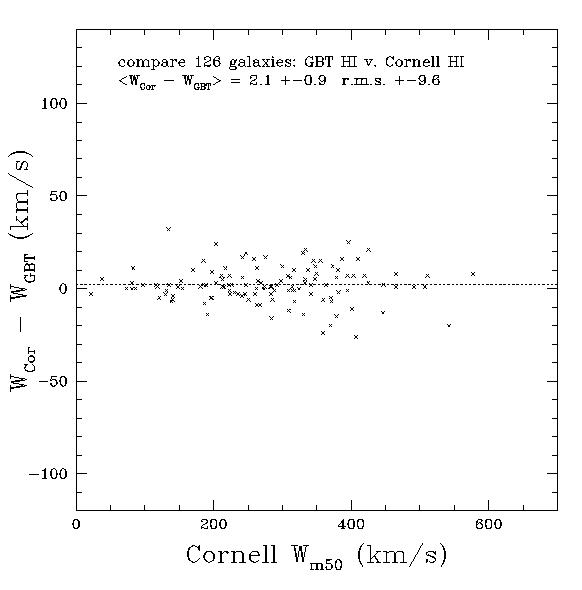}
\plotone{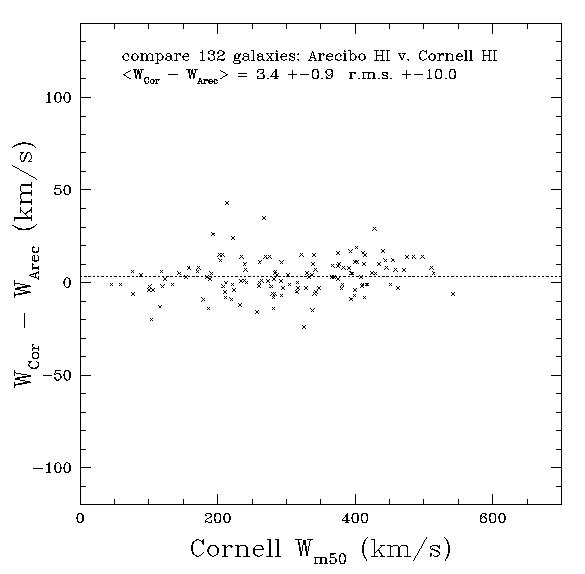}
\caption{Comparison between our new data and data from the Cornell archive, both analyzed with the same pipeline designed to measure the linewidth at 50\% of mean flux.  Top: New data acquired with the Byrd Green Bank Telescope.  Bottom: New Data acquired with the Arecibo Telescope.}
\label{wcorn-wnew}
\end{figure}

\section{Comparison of New and Old Data}

The comparisons in the previous section are between different measures of linewidth with largely the same data.  In this section there is a comparison of new and old observations of identical targets, all analyzed with the procedure discussed in this paper.   A point of detail: the linewidths discussed in this section have received the small correction for spectral resolution discussed in the next section to facilitate the comparison of observations made with different telescopes and receivers.
Figure~\ref{wcorn-wnew} shows the current status of comparisons between new and Cornell archival profiles, giving separate consideration to new GBT and Arecibo observations.  The archival profiles come from a variety of telescopes, never GBT.  In all cases, the linewidth measure is our $W_{m50}$ parameter.

The r.m.s. scatter between linewidths drawn from material out of the Cornell HI archive and linewidths determined from new observations is a satisfactory 10~\kms.  If errors are partitioned equally, the implied uncertainty is 7~\kms\ in each of the new and old measures.  The scatter is similar in the separate GBT and Arecibo comparisons.

There is a minor mystery in the zero-point offsets seen in both samples displayed in Fig.~\ref{wcorn-wnew}.  The linewidths measured from the archival profiles are 3~\kms\ wider in the mean, a difference with $3 \sigma$ significance.  It may be a factor that the new profiles have higher signal-to-noise in the great majority of cases.  This is not a negative reflection on the archival material; simply a consequence of our strategy of primarily re-observing objects with poor profiles (though recalling that profiles ascribed errors larger than 20~\kms\ are rejected in all comparison samples).  With the 258 galaxies represented in Fig.~\ref{wcorn-wnew}, the mean difference in the error assigned to the linewidth, archive minus new, is 6~\kms.  An explanation for the zero-point offset $might$ be that linewidths are slightly overestimated with noisier profiles.   

\begin{figure}[htb!]
\figurenum{19}
\plotone{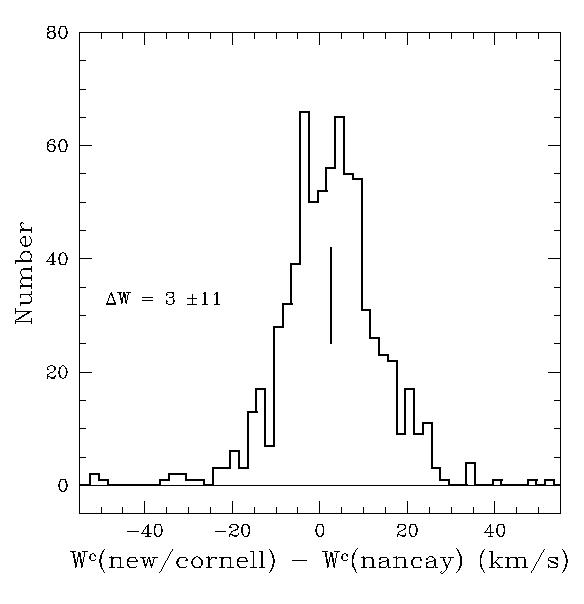}
\caption{The average of our new linewidths and linewidths derived from the Cornell archive data compared with linewidths derived from data from the Nan\c cay archive.  All data are analyzed with the same pipeline designed to measure the linewidth at 50\% of mean flux.}  
\label{wother-wnan}
\end{figure}

Another large component of the {\it All Digital HI} catalog is built from data extracted from the Nan\c cay database.  To be clear, the profile fits and derivative parameters for observations from the Nan\c cay telescope given through the {\it All Digital HI} catalog are based on the analysis procedures described in this paper; results from the original source of the data are found in the catalog {\it HI Nancay}.  This is analogous to the distinction in the case of the material from Cornell, with original source material in catalog {\it Springob/Cornell HI} and re-analyzed material in catalog {\it All Digital HI}.   There are 720 galaxies in {\it All Digital HI} with both a satisfactory ($e_W \le 20$~\kms) Nan\c cay linewidth and a satisfactory linewidth either from the Cornell archive or new as reported here.  The difference between new/Cornell and Nan\c cay widths is $<W_{new/cornell} - W_{nancay}> = 2.6 \pm 0.4$ with r.m.s. scatter 10.8~\kms\ after rejection of 7 cases with excursions in excess of $4 \sigma$.
The comparison is shown graphically in Figure~\ref{wother-wnan}.  The r.m.s. scatter is at a level that, in comparison with the discussion surrounding Fig.~\ref{wcorn-wnew}, implies a characteristic uncertainty with the Nan\c cay widths of 8~\kms.

The offset of 2.6~\kms\ seen in Fig.~\ref{wother-wnan} is small but statistically significant.  The comparison shown here is with widths corrected for line broadening.  If the comparison is made on the directly observed $W_{m50}$ linewidths then the difference is $-0.1$~\kms.  The offset arises through the broadening corrections discussed in the next section.  The Nan\c cay data tends to need larger corrections and our recipe may be slightly excessive for this sample.  However, the problem is sufficiently small that it will not affect the measurement of distances.

\begin{figure}[htb!]
\figurenum{20}
\plotone{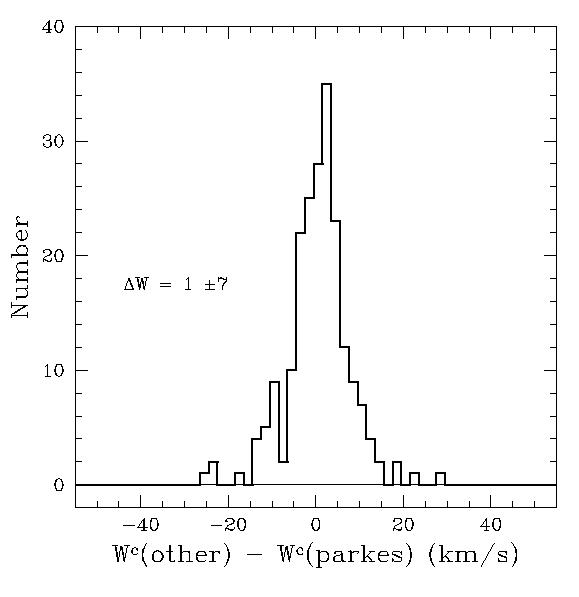}
\caption{The average of our new linewidths and linewidths derived from the Cornell and Nan\c cay  archive data compared with linewidths derived from data from the Parkes archive.  All data are analyzed with the same pipeline designed to measure the linewidth at 50\% of mean flux.}  
\label{wother-wpaks}
\end{figure}

At the most southerly latitudes, neutral hydrogen observations require use of the Parkes Telescope and results are becoming available on their archive\footnote{http://www.atnf.csiro.au/research/multibeam}.
An important contribution has come from HIPASS, the HI Parkes All-Sky Survey  \citep{2004AJ....128...16K}.   A comparison with material drawn from this database is summarized in the histogram of Figure~\ref{wother-wpaks}.  The difference in linewidths measured off Parkes spectra with those measured off spectra from the Cornell or Nan\c cay archives or from own new observations for 205 galaxies is $<W^c_{other} - W^c_{parkes}> = 0.7 \pm 0.5$ with r.m.s. scatter 7.3~\kms.  The excellent very low scatter can be attributed in part to the fact that the galaxies in the comparison tend to be nearby and easily detected in HI.

\section{Linewidth Adjustments}

Several systematics affect linewidths.  Two that are well understood are a slight relativistic broadening and broadening because of finite spectral resolution.  SHGK discuss these matters at length and we adopt a simplified version of their solution.  We adjust linewidths with the equation:
\begin{equation}
W_{m50}^c = {{W_{m50}} \over {1+z}} - 2 \Delta v \lambda
\label{Wc}
\end{equation}
where $cz$ is the heliocentric velocity of the galaxy, $\Delta v$ is the spectral resolution after smoothing, and $\lambda$ is determined empirically.  SHGK, with a closely equivalent formula, give a convoluted recipe for $\lambda$.  Their complex description may be appropriate in their case because of a coupling between signal-to-noise and their peak measurement, hence their linewidth measure.  With our parameter based on mean flux there is less systematic dependence on signal-to-noise.  From tests on profiles with successively increased smoothing, we find broadening is statistically described by Eq.~\ref{Wc} if $\lambda = 0.25$.  This correction is close to those advocated by \citet{1990A&AS...82..391B} and \citet{2001A&A...370..765V}.

The measured profile linewidth, whether it is our $W_{m50}$ or another, is only a parameter of observational convenience and it is desirable to translate it into something more physically meaningful.  Rectification to edge-on orientation is standard (division by sin~$i$ where $i$ is the inclination from face-on).  It is also common, but less secure, to adjust the observed linewidth to correspond statistically with twice the maximum rotation velocity, $V_{max}$.  The adjustment is based upon samples with both global profiles and detailed rotation curves.  \citet{1985ApJS...58...67T} investigated the matter and provided a description that accounts for the effects of broadening by turbulent motions that transitions from a linear to quadratic correction as the unbroadened profile transitions from roughly boxcar in giant galaxies to gaussian in dwarfs.  Their formula is:
\begin{eqnarray}
\nonumber
W_{R,\ell}^2  = W_{\ell}^2 + W_{t,\ell}^2 [1 - 2  e^{-(W_{\ell}/W_{c,\ell})^2}] \\
 -  2 W_{\ell} W_{t,\ell} [ 1 - e^{-(W_{\ell}/W_{c,\ell})^2}]  
\label{WR}
\end{eqnarray}
where the subscript $\ell$ stands for the observed linewidth measure,  $W_{t,\ell}$ is the turbulent broadening for that observed measure, and $W_{c,\ell}$ characterizes the transition from boxcar to gaussian intrinsic profiles.  In the case $W_{\ell} = W_{20}$, \citet{1985ApJS...58...67T} recommended $W_t =38$~\kms\ and $W_c = 120$~\kms.

More recently, the problem has been studied in detail by \citet{2001A&A...370..765V}.  They compared global profiles with detailed rotation curve information for galaxies observed with the Westerbork Synthesis Radio Telescope.  They determined that the Tully-Fouqu\'e value taken for $W_t$ was too large; that to get $<W_R - 2 V_{max} ~{\rm sin}~i> \simeq 0$ with the transformation of $W_{20}$ to $W_R$ of Eq.~\ref{WR} requires $W_t = 22$~\kms.

\begin{figure}[htb!]
\figurenum{21}
\plotone{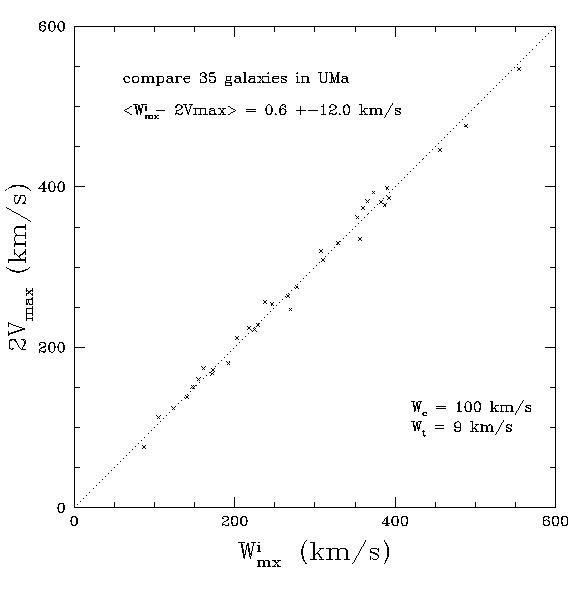}
\caption{Comparison between the global profile parameter $W_{mx}^i$ and twice the maximum rotation velocity determined from spatially resolved rotation curves. 
}
\label{wmi_2vmax}
\end{figure}

The transformations can be expected to be slightly different in detail with the new linewidth parameter $W_{m50}$.  Comparisons have been made with 35 galaxies in the Ursa Major Cluster with rotation curves determined from observations with the Westerbork Synthesis Radio Telescope and $V_{max}$ values reported by  \citet{2001ApJ...563..694V}.  As anticipated by \citet{2001A&A...370..765V}, the effect of measuring the linewidth at a higher flux level above the baseline requires reduction of $W_{c,\ell}$ and, especially, $W_{t,\ell}$.  The optimal fit results in the correlation seen in Figure~\ref{wmi_2vmax}.  To differentiate from parameter variations discussed in earlier publications we define $W_{mx}^i \equiv W_{R,m50}^i$ and find a best fit for the parameters in Eq.~\ref{WR} with $W_{c,m50} = 100$~\kms\ and $W_{t,m50} = 9$~\kms.  With these parameters, observed $W_{m50}$ linewidths are transformed into $W_{mx}^i$ linewidths that agree with $2 V_{max}$ with an r.m.s. scatter of 12~\kms\ after deprojection (r.m.s. scatter 10~\kms\ in the line-of-sight).  This scatter is comparable to the $W_{m50}$ measurement accuracy.

A detailed study of the relationship between observed linewidths and the intrinsic kinematics of galaxies was carried out by \citet{2008PhDT.........3S}.   Such a study is particularly important if the interest is to understand the physical basis for the relationship between galaxy rotation and the light distribution.  The slope of the correlation can be affected by the details of measurements and adjustments.  For the practical matter of measuring distances the greater importance is to be consistent.

\section{The All Digital HI  Catalog in EDD}

\begin{figure}[htb!]
\figurenum{22}
\plotone{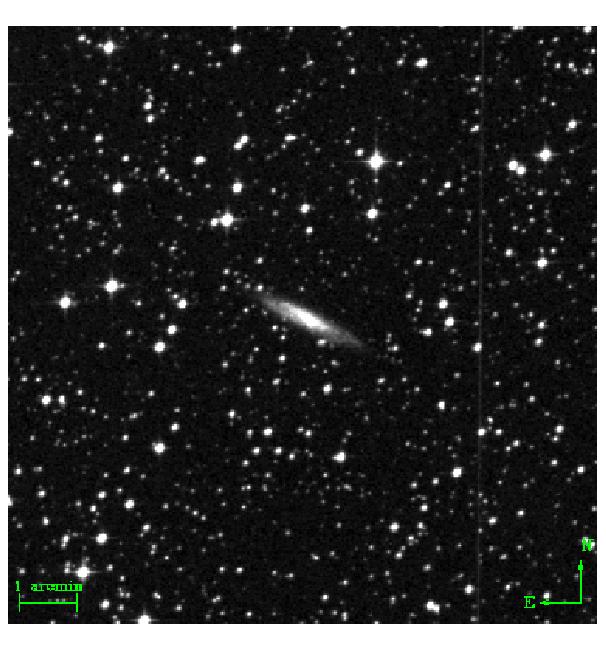}
\plotone{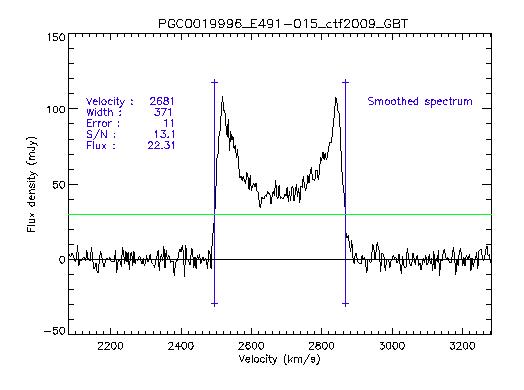}
\plotone{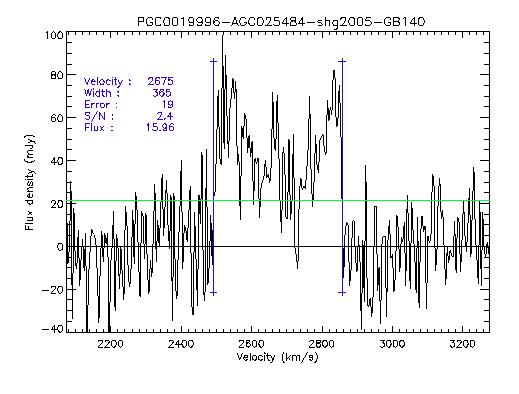}
\caption{Example of a graphical display accessed from the {\it All Digital HI} catalog.  
}
\label{PGC0019996}
\end{figure}

The {\it All Digital HI} catalog is accessed by selecting the `next' button on the EDD home page.  It can be selected alone or in tandem with any of the other catalogs and either all or any fraction of the elements within the catalog can be selected.  The tabular portion of the catalog is displayed with the `select' button and can be exported with the `download' button.  Upon entering the tabular display, one can navigate to graphical displays of HI profiles by selecting on the common name of a galaxy (selecting on the PGC name in this, and any of the other catalogs, brings up a digital sky survey image of the galaxy).  An example of what will be found is shown in Figure~\ref{PGC0019996}.  The galaxy seen in this case is PGC 19996 = ESO 491-015.  The image of the galaxy is drawn from the LEDA website and displayed with a field of 10 arcmin, roughly the beam size of the GBT, hence a scale reasonable for an inspection for contamination from any near neighbors.  The left profile was acquired by the authors with observations using GBT.  The right profile is based on observations with the Green Bank 140-foot telescope and was obtained from the Cornell archive.  The two profiles are shown after treatment by the same analysis pipeline.

The results of the analysis are carried to Table 4, the main catalog,\footnote{Table 4 is provided with the on line version of the article in the Astronomical Journal or a continuously updated version can be downloaded from EDD at http://edd.ifa.hawaii.edu upon selecting the {\it All Digital HI} catalog.} which includes the following information.  The parameters given in columns 3--6 are averaged over multiple observations.  Those in columns 7--18 are for an individual observation and the columns repeat if there are multiple observations.  The first few lines of the table are given in the print version of this article.

1. Principal Galaxies Catalog (PGC) name from the Lyon Extragalactic Database (LEDA).

2. Common name (click on name to view profiles).

3. Weighted average heliocentric velocity from all acceptable profiles (\kms).  Weights are based on the inverse square of assigned errors.

4. Weighted average linewidth approximating twice the maximum rotation velocity before projection from all acceptable profiles, $W_{mx}$ (\kms).

5. Uncertainty attributed to the linewidth, the inverse square root of the sum of individual weights (\kms). 

6. Number of acceptable profiles (errors less than or equal to 20~\kms).

7. Source of observation.  Source codes are identified in Table 3 for sources incorporated at the time of publication.

8. Telescope and receiver. See Table 2 for more information.

9. Heliocentric velocity, the midpoint between the low and high velocities that define $W_{m50}$ (\kms).

10. $W_{m50}$: linewidth at 50\% of the mean flux per channel where the mean flux is calculated within the 90\% window, the range of velocities excluding 5\% of the integrated flux at each end of the profile (\kms).

11. Linewidth corrected for relativistic and instrumental broadening, $W_{m50}^c = {{W_{m50}}\over{1+z}} - 2 \Delta v \lambda $ where $\lambda = 0.25$ and $\Delta v$ is the product of the values in columns 16 and 17.

12. Linewidth adjusted to statistically equal twice the maximum rotation velocity, before deprojection, derived from spatially resolved rotation curves (\kms).  Statistically,  $W_{mx} \sim 2 V_{max} {\rm sin} i$.    This parameter is only considered meaningful if  $e_W \le 20$~\kms.

13. Uncertainty in the linewidth (\kms).  Uncertainties less than or equal to 20~\kms\ are considered adequate for the purpose of determining a distance through the luminosity--linewidth correlation.  An initial assignment of uncertainty, $e_W$, is based on signal-to-noise ($S/N$):   $e_W = 8$~\kms\ if $S/N \ge 17$;  $e_W = 21.6 - 0.8 S/N$ if $17 > S/N > 2$; $e_W = 70 - 25 S/N$ if $S/N < 2$.  If the spectral resolution after smoothing is greater than this assignment then the error  is increased to match the smoothed resolution.  The uncertainty may have been modified by manual intervention to either increase it to above 20~\kms\ if the profile is too poor to be used for a distance measure or to decrease it to equal or below 20~\kms\ if the profile is considered adequate for this purpose.  The error value $e_W = 100$ \kms\ is used to signal cases of confusion and the error value $e_W = 500$ \kms\ identifies a dubious/null detection.

14. Signal-to-noise ($S/N$).  The signal is the mean flux per channel within the velocity range of the 90\% window.  The noise is calculated over 100 channels on each side of the signal outside the velocity range of the 100\% window.

15. The flux is the signal integrated over all channels within the 100\% window (Jy \kms).  No attempt has been made to account for flux lost due to the finite beam size.

16. Channel resolution (\kms).

17. An integer $N$ indicates averaging over $N$ spectral channels in the profile that is displayed.

18. $F_{m50}$ is the flux level at 50\% of the mean, the level at which the measurement of $W_{m50}$ is made.  

The most recent status of the full Table 4 catalog is made available with the electronic version of this article.  

\section{Summary}

The purposes of the {\it All Digital HI} catalog in EDD are threefold: first, to make available the results of new observations made of galaxies in the 21cm HI line; second, to make it easy to compare results with other observations and link to other information about the targets;  third, to present a reanalysis of all archival data available in digital form to ensure that consistent linewidth information is available for essentially all galaxies that have been observed in HI.

Our preferred linewidth parameter is $W_{m50}$, the profile width at 50\% of the mean flux within the velocity window containing 90\% of the total flux.  This parameter is a variant of one of those introduced by the Cornell group \citep{2005ApJS..160..149S}.  The availability within EDD of several very large catalogs of HI information facilitates comparisons and provides a way of culling bad data.  It is satisfying to see the tight correlations between alternative profile descriptors.  It is to be appreciated that various alternative profile descriptors have merit but it is important for the measurement of distances to maintain consistency.  After accounting for zero point offsets, r.m.s. scatter between alternatives is at the level of 10~\kms.  Based on comparisons with detailed rotation curve information, a statistical transformation is proposed that takes the observed global linewidths to an approximation of the maximum rotation velocity $V_{max}$.  

Presently the catalog {\it All Digital HI} contains 15,411 profiles providing information on 13,423 galaxies.  Some 57\% of the profiles originate from the Cornell database (8740), 21\% originate from the Nan\c cay database (3225), 7\% come from Arecibo Legacy Fast Alfa survey (1047), another 7\% were extracted from the Parkes archive (997), 1\% are an Effelsberg contribution independent of the Cornell database (176), and 8\% result from new observations by our collaboration (1225).  Currently profiles for 10,580 galaxies are deemed acceptable.  In 1,330 cases there are at least two acceptable profiles and in 81 cases there are three acceptable profiles.  Inter comparisons between sources suggest that the characteristic accuracy of an individual acceptable profile width is 7~\kms.

\acknowledgements{
New observations across the entire sky have been made possible by access to three fine radio telescopes.  We made early observations with the refurbished Arecibo Telescope and expect to add fresh material coming from the wide field multi-beam survey.  At the Green Bank Telescope our ongoing project Cosmic Flows has been awarded the status of a Large Program.    Observations of the deep southern sky began in 2009 with the Parkes Telescope in Australia.   Equally important to us has been access to archival material from the Cornell Digital HI Archive, the Nan\c cay Radio Telescope HI profiles of Galaxies database, and the Australia Telescope online archive.  Although electronic archives are a great innovation, the low-tech information gathered in the {\it Pre Digital HI} catalog retains great value and we thank Cyrus Hall for his role in assembling that material.   We have made extensive use of NED, the NASA/IPAC Extragalactic Database operated by the Jet Propulsion Laboratory, California Institute of Technology, and the HyperLeda database hosted at the Universit\'e Lyon 1.  Web access to the {\it All Digital HI} catalog is found at http://edd.ifa.hawaii.edu.
}

\bibliography{paper}
\bibliographystyle{apj}

\clearpage

\end{document}